\documentclass[a4paper,10pt,twocolumn]{article}
\usepackage[utf8]{inputenc}
\usepackage{lipsum}
\usepackage{moreverb,url}
\usepackage{amssymb}
\usepackage{amsthm}
\usepackage[english]{babel}
\usepackage{csquotes}
\usepackage{lipsum}
\usepackage{subfigure}
\usepackage{amsmath}
\usepackage{esint}
\usepackage{amsfonts}
\usepackage{graphicx}
\usepackage{lmodern}
\usepackage{ragged2e}
\usepackage{pdflscape}
\usepackage{rotating}
\usepackage{multirow}
\usepackage{xcolor}
\usepackage{biblatex}
\usepackage{epstopdf}
\usepackage[a4paper, hmargin={2cm, 2cm}, vmargin={2cm, 2cm}]{geometry}
\graphicspath{ {./figs/} }

\title{Aerodynamic Influence Over Leading and Pursuing Motorcycles Equipped With Downforce-Generation Wings}

\author{Braulio Gutierrez Pimenta\footnote{Corresponding author, e-mail: braulio.pimenta@gmail.com} , Luís Paulo de Queiroz Moreira,\\ Adriano Possebon Rosa and Roberto Francisco Bobenrieth Miserda. \\ Mechanical Engineering Department, University of Brasília.}

\begin{document}

\maketitle

\begin{abstract}
The aerodynamic influence of a wing-equipped motorcycle on a pursuing motorcycle presents critical implications for stability and performance. This study investigates the induced flow dynamics, characterized by a turbulent and complex wake that significantly affects the aerodynamic forces and moments experienced by the following motorcycle. The presence of aerodynamic appendices on the leading motorcycle intensifies these effects, generating coherent wingtip vortices that propagate downstream-a typical behavior of lift-generating devices. In this work, numerical simulations reveal that the aerodynamic consequences vary with the relative positioning of the pursuing motorcycle. A lateral offset can reduce wheelie tendencies due to beneficial flow interactions, while lateral alignment and longitudinal positioning variation may exacerbate negative aerodynamic impacts, compromising stability. Contrary to initial expectations, the simulations in this work demonstrate that the turbulent wake and the coherent vortex pair influence the pursuing motorcycle's behavior independently, persisting across all tested relative distances. While the turbulent wake usually creates a low-pressure region that facilitates drafting, the presence of coherent vortices reduces this advantage by introducing additional lift through the upwash velocity component. Conversely, specific lateral deviations can lessen wheelie effects, with downwash becoming the dominant flow component. Although difficult to be removed promptly due to being part of the overall high performance motorcycle design, it is suggested that the downforce generating aerodynamic appendices removal should be considered by the motorcycling competition governing bodies to provide better safety and racing conditions at all categories that make use of it.
\end{abstract}

\noindent Keywords: Motorcycle Racing, Aerodynamics, Wingtip Vortex, CFD, RANS, OpenFOAM.


\section{Introduction}
Motorsport has been the main laboratory for developing technologies for the road mobility consumer market. The eagerness to win races brought improvements in performance, safety and also riding comfort. Since their production and development at the beginning of the 20th century, cars and motorcycles brought these improvements through the dawn of motorsport. A rapid increase in velocity especially in motorcycles showed that safety issues had to be addressed. Several areas of development had to deal with safety along with performance.


As an example, one of the first road races to be held was the Tourist Trophy (TT), located in the Isle of Man, Northern Ireland. The first TT was held in 1907 \cite{iomtt_url_history} and in 1920 the lap mean speed record was of 55.62 mph (89 km/h) on the Snaefell Mountain Course. One hundred years later the race is still held in roughly the same road circuit, where the current speed record is of 136.358 mph (219.447 km/h) by Peter Hickman in 2018 in the Superbike category \cite{iomtt_url_speed_records}. The unofficial maximum speed is of 206 mph (332 km/h), held by Bruce Anstey in 2006. What followed in the subsequent decades was the professional championships taking place, organized by their respective Federations, such as the FIM Road Racing World Championship Grand Prix, organized by the \emph{Fédération Internationale de Motocyclisme} (FIM) in 1949. Nowadays, the main motorcycle championship is the MotoGP, also organized by FIM. The MotoGP class motorcycles are prototypes within a maximum four cylinder fuel injected engine capacity of 1,000cc. These engines may produce up to 300 bhp, resulting on a maximum speed of 366.1 km/h, held by Brad Binder in the Mugello circuit in 2023 \cite{motogp_speed_record}. Other professional championships include the FIM Superbike World Championship (WSBK), also regulated by FIM, where street legal motorcycles are modified to comply with a simpler set of rules than those of MotoGP prototypes. The WSBK performance almost equals their MotoGP counterparts, where production based motorcycles race with minimal chassis and engine modifications.


Other areas of development include braking systems with heat-resistant materials, such as carbon braking discs and radial-mount calipers. Electronic riding ads, for example, the anti-lock braking system (ABS), traction control (TS) and anti-wheeling systems. Motorcycle suspension improvements are the inverted telescopic suspension forks, holeshot and height adjustment devices. The racing tire also takes place with specific compounds, thus providing extra grip at a narrow range of temperatures. The chassis design is also taken into account for controlled rigidity during the braking and turning stages \cite{cossalter_2006}.

\begin{figure}[ht!]
\centering
\includegraphics[width=\columnwidth]{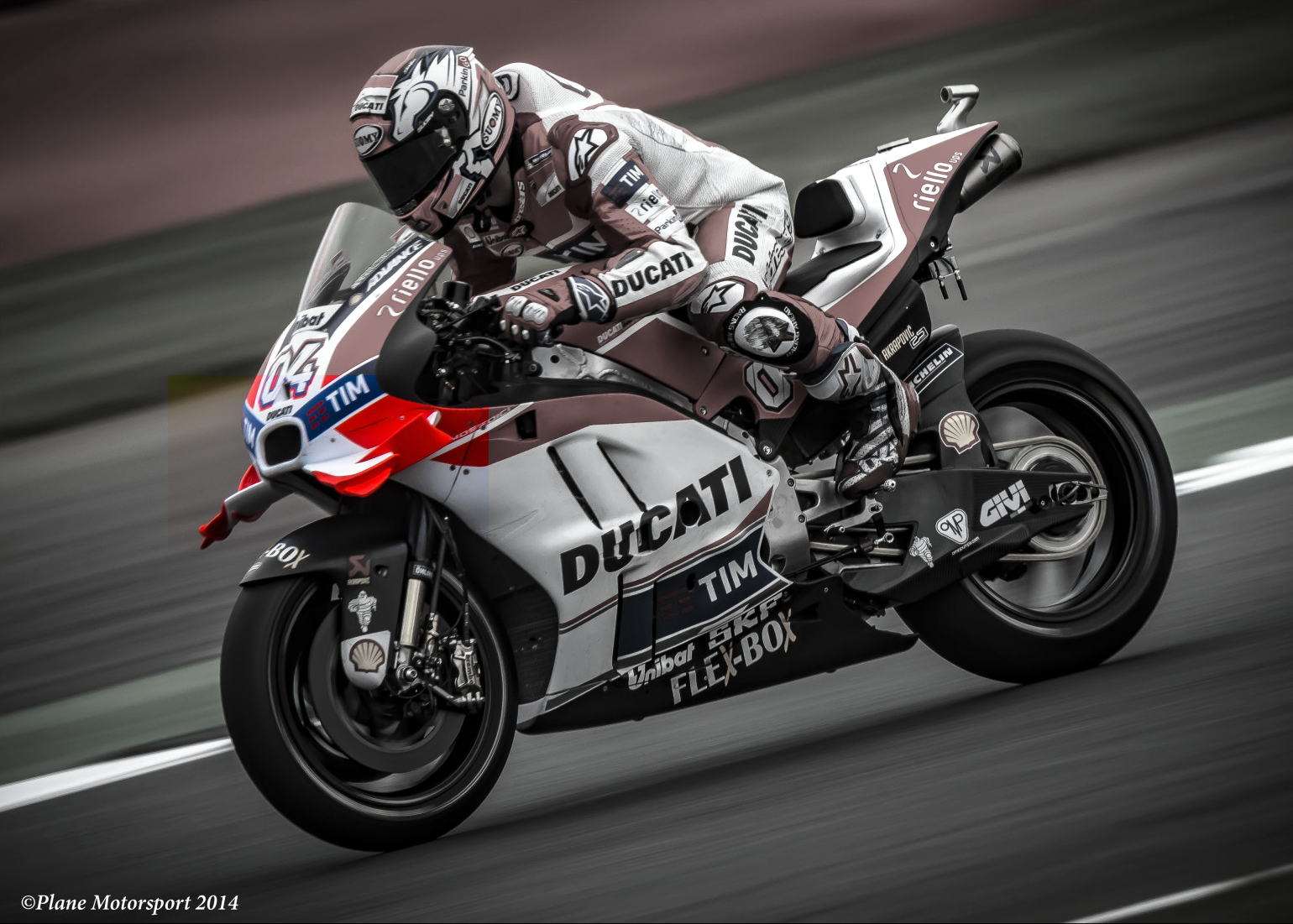}
\caption{Andrea Dovizioso riding the wing-fitted Ducati MotoGP motorcycle during the 2016 season \cite{ducati_2016} (wing colorized in detail).}
\label{fig_01}
\end{figure}

Maximum speed grew steadily during the post second world war era with better materials and engine development along with further improvement in other areas. The use of aerodynamic fairings began to take place to alleviate the undesirable aerodynamic effects and to put the rider into better riding conditions. The primary effect sought by motorcycle fairing placement is the drag reduction, thus resulting also into higher acceleration. The use of wings to generate downforce has been introduced in modern racing by the Ducati Corse team in 2016, where simple wings were fitted into the front fairing, as can be seen in figure \ref{fig_01}. This was the first time that wings in the front section of a racing motorcycle were taken as a permanent feature and dictated by the sports rules, being still used in the 2024 MotoGP season \cite{motogp_rules_2024} and that will be allowed in the overhaul rule changes of 2027 onwards, where front wings must comply with smaller dimensions along with smaller engine size with a displacement of 850 cubic cm and cylinder bore of 75 mm to reduce extreme revving engines \cite{motogp_2017_rules}. Front wings in motorcycles helps performance in three main stages of racing:

\begin{itemize}
    \item Braking stage: most of the braking force is due to the front tire grip, where weight is transferred to the front. The front wing increases the tire static friction, allowing the motorcycle to be decelerated at a higher rate.
    \item Corner entering stage: this stage is referred to as before the motorcycle reaches the corner apex. By positioning front wings with a negative dihedral angle and reducing the turning side wing effectiveness with rider body positioning, the outside wing generates a downforce component even with a higher motorcycle camber angle, thus allowing more aggressive trail braking, consequently reducing deceleration time before hitting the corner apex. In the modern motorcycle racing era, the corner entrance stage is the most common phase where accidents occur, where the front wheel loses traction.
    \item Corner exit stage: upon exiting corners after passing the apex, the racing motorcycle must attain maximum acceleration. This causes the wheeling effect when the motorcycle is at a more upright position, when the front wheel achieves zero load \cite{cossalter_2006}. This can be avoided with anti wheeling electronic systems by reducing engine torque when the chassis is at a certain pitch angle. The other way to reduce wheeling is the use of front wings that produces downforce, resulting on a zero pitching moment for optimum acceleration. Since the latter does not reduce the engine power output, it is usually the better option, albeit generating more aerodynamic induced drag. Thanks to electronic aids, such as traction control, uncontrollable wheel spin may be prevented, making highsides and lowsides followed by accidents much rarer nowadays at the corner exit stage.
\end{itemize}


The aforementioned effects are related to a single motorcycle, where no secondary aerodynamic disturbances from other sources are regarded, for example, the aerodynamic disturbances originating from the wake of leading motorcycles. When taken into account, these disturbances may also be beneficial for trailing motorcycles in the performance sense, such as the drafting effect providing overall drag reduction, coming mainly from the pressure drag reduction due to flow separation (form drag), since motorcycles are naturally blunt bodies, even on fairing equipped ones.

Unfortunately the benefits of downforce produced by front wings may be reduced when the turbulent low-pressure wake impinges over the front wings, reducing its downforce and thus the effects listed above, providing a dangerous performing condition for the trailing motorcycle when not under nominal operating conditions. A fine example is in the final race of the 2023 MotoGP season at Valencia, when the trailing Jorge Martin could not brake at the same performance level as Francesco Bagnaia at the end of the main straight\footnote{\url{https://www.youtube.com/watch?v=ZK-HZIUEe84&t=80s}}, even though by having the same tire allocation \cite{motogp_valencia_2023} and by both pilots riding same motorcycle model. It can be noted, however, that other issues may have influenced the outcome at some degree, such as motorcycle adjustment and tire degradation, but the lack of aerodynamic downforce may also be regarded as one of the main causes of the near crash incident.

Little is publicly available when motorcycle aerodynamics is concerned. Since it is regarded as sensitive knowledge applicable in motorsport racing, only a few academic works may be found in literature, where CFD simulations are the norm while wind tunnel testing constitute rare appearances due to its high cost and model complexity. Some of these works include the work of Peri and Capuana \cite{peri_masters_2023}, although mostly dedicated to design and selection of wings for motorcycles, presents OpenFOAM simulations of motorcycles with and without wings in straight movement and in cornering situations. The work of Blocken et al. \cite{blocken_2020} analyzes the aerodynamic effects of the vortex wake vortices generated by a motorcycle on a bicycle, without lateral distance but varying the longitudinal distance. The simulations were carried out in Ansys Fluent. The work of Wi\'nski and Piechna \cite{winski_2022} addresses the computational simulation of the flow aerodynamic development around a single motorcycle, under a variety of approaches. One Interesting conclusion comes from the study of RANS turbulence models, where the SST-DDES and SST-SAS models presents more accurate results at a higher computational cost. With this in mind, they have found that the model k-$\omega$-SST proved to be one of the most accurate among the RANS models discussed. Another CFD work was carried out by Fintelman et al. \cite{fintelman_2015}, where the aerodynamic forces are evaluated under a motorcycle subject to crosswinds to assess its safety margins. They used the SST-DDES and RANS turbulence models where used in their simulations. Other works include wind tunnel and CFD comparative studies for performance \cite{araki_2001}, comfort improvement \cite{takahashi_2009} and basic flow characteristics \cite{watanabe_2003}. In our literature search, we found no literature paper that investigates the effects of the wing on the pursuing motorbike, while most of the studies found focus only on the aerodynamic influence of undisturbed flow.

This work is a numerical study of the aerodynamic influence over a trailing motorcycle when front wings are considered. All the conditions of wing-equipped and unequipped motorcycles are considered on a wide range of longitudinal and lateral distances for aerodynamic influence assessment. This work aims a better understanding of rider safety under racing conditions where aerodynamics plays a major role on a competition naturally dangerous, where a noticeable increase in the number of accidents is taking place each year \cite{motogp_dangerous_2023}.

This work is divided into the following sections: aerodynamic theory, where the physical basis used to corroborate the hypothesis to be evaluated in the results is presented. The numerical setup section presents the details of the numerical method and the modeling and discretization details. The results section presents the simulation results for several cases: with-fitted and plain fairing are considered for trailing and leading motorcycles at varying longitudinal and lateral positions. Finally, the conclusion is presented where a final assessment of wing-fitted motorcycles safety is presented.

\section{Aerodynamic Theory}
A basic sports bike fairing is comprised of a smooth light structure placed in the front section of the motorcycle in order to deflect the incoming air, while the pilot is tucked inside to reduce the frontal area during the accelerating phase. This has been the basic appendages used in racing motorcycles and their primary use is to smooth out the air flow throughout the motorcycle and reduce its aerodynamic drag. Further improvements are mostly related to its size and shape.

\begin{figure}[ht!]
\centering
\includegraphics[width=0.49\columnwidth]{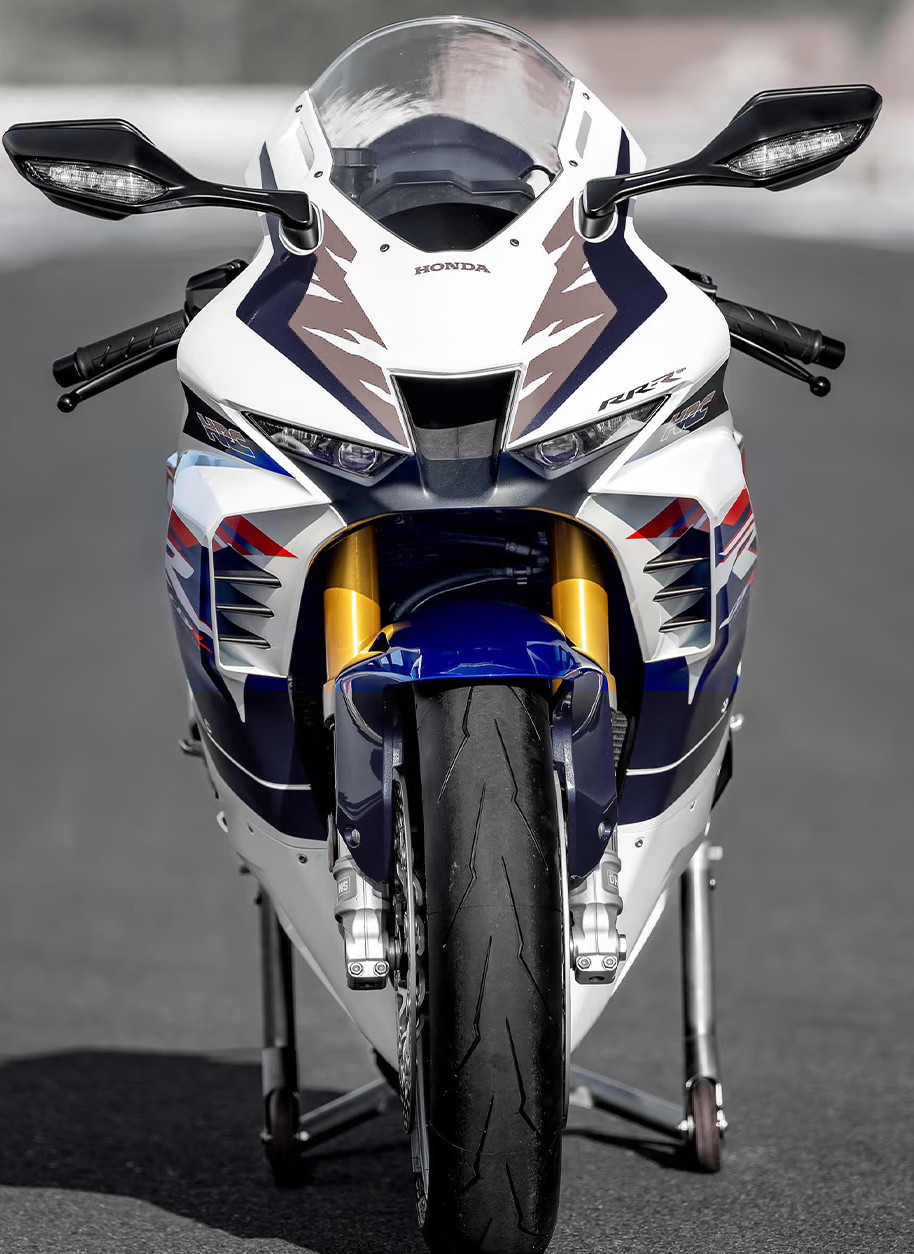}
\includegraphics[width=0.49\columnwidth]{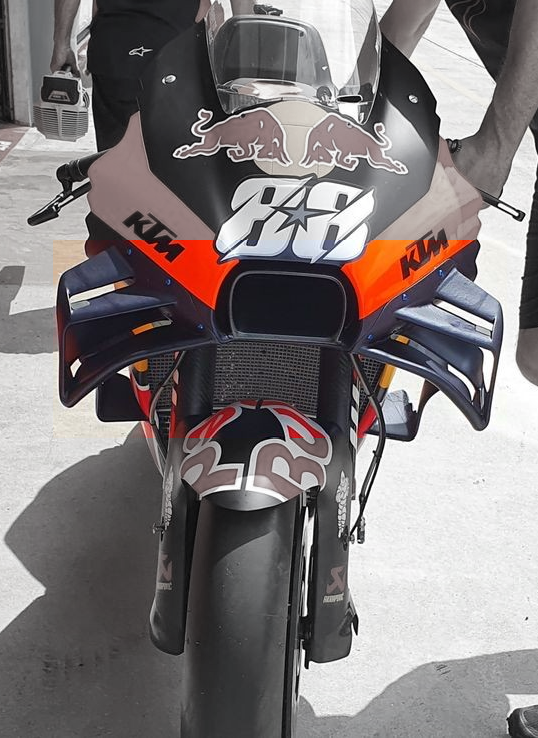}
\caption{Front diffuser fairing of the street legal Honda CBR 1000RR-R (left) \cite{cbr1000rr} and Miguel Oliveira's KTM RC16 2022 MotoGP prototype \cite{kmt_motogp_2022} with front wing mounted fairing (right).}
\label{fig_03}
\end{figure}

More recently front wings were introduced to generate downforce on the front tire and thus increase performance as mentioned before. Due to its overall shape, sporting motorcycles equipped with fairings also generate a lift force \cite{cossalter_2006}, decreasing the front tire load and reducing the motorcycle's stability. Two types of aerodynamic appendages may be used to reduce this negative effect: front wings and diffusers. Figure \ref{fig_03} shows two examples of wing-mounted and diffuser-mounted motorcycles. These two devices serve the same purpose of generating downforce, while their aerodynamic effects may show differing features, such as flow wake and coherent flow structures. This work shall consider only the wing-mounted motorcycles since their use has been widespread in motorcycle motorsport, while the use of diffusers is limited use to only a couple of manufacturers.

Even racing motorcycles mounted with fairings cannot be seen as a pure aerodynamic shape, while a large turbulent and chaotic wake follows the motorcycle and rider. This wake has the same size as the motorcycle height where its turbulent kinetic energy comes from the engine power being dissipated as drag. This drag is characterized as profile drag, where the front is under large pressure and the aft region is under low-pressure. The adverse pressure gradient along the boundary layer over the motorcycle and rider's surface produces recirculating regions near the body that are then convected along with the flow as a wake. This wake has a low-pressure region near the leading body that is used for drafting by a pursuing motorcycle. Since boundary layer detachment is distributed along a large area of the motorcycle and rider, the dominant drag comes from pressure distribution and thus shear drag from air viscosity on blunt bodies can be regarded as negligible \cite{anderson_2017,batchelor_2000}. Suitable modeling and calculation of the turbulent wake must be taken into account to predict reliable results.

Typical velocities in motorcycle racing ranges from 100 km/h (27.7 m/s) to 350 km/h (97.2 m/s), resulting into a Reynolds number ranging from  $2.74\cdot10^6$ to $9.61\cdot10^6$ when typical air conditions are considered; reference kinematic viscosity of $\nu_{\infty}=1.436\cdot10^{-5}$ m$^2$/s and for a sports bike the typical characteristic length is the wheelbase of $L_{\text{ref}}=1.42$ m. Turbulence takes an important role in this flow regime \cite{boundary_layer_2017} and its transport phenomena must be taken into account for the flow modeling.

The aforementioned flow features are based on a typical motorcycle with no downforce-generating devices, where a turbulent wake is mostly generated by the blunt body format. Front wing-fitted motorcycles, when subjected to air flow, will have their front wings generating the wing tip vortexes and the induced drag is one of the issues related to lift generation.


\begin{figure}[ht!]
\centering
\includegraphics[width=0.48\columnwidth]{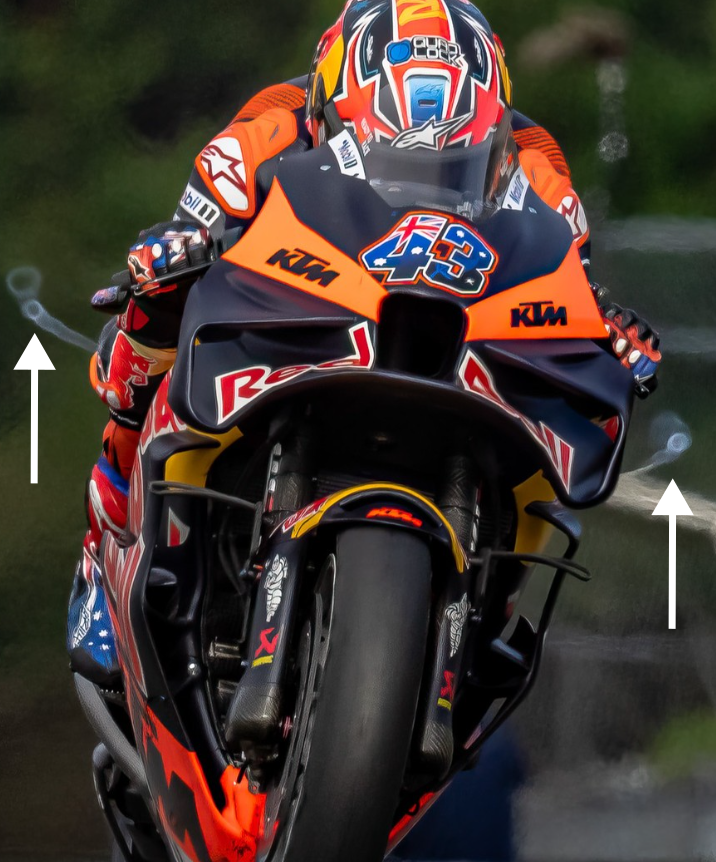}
\includegraphics[width=0.48\columnwidth]{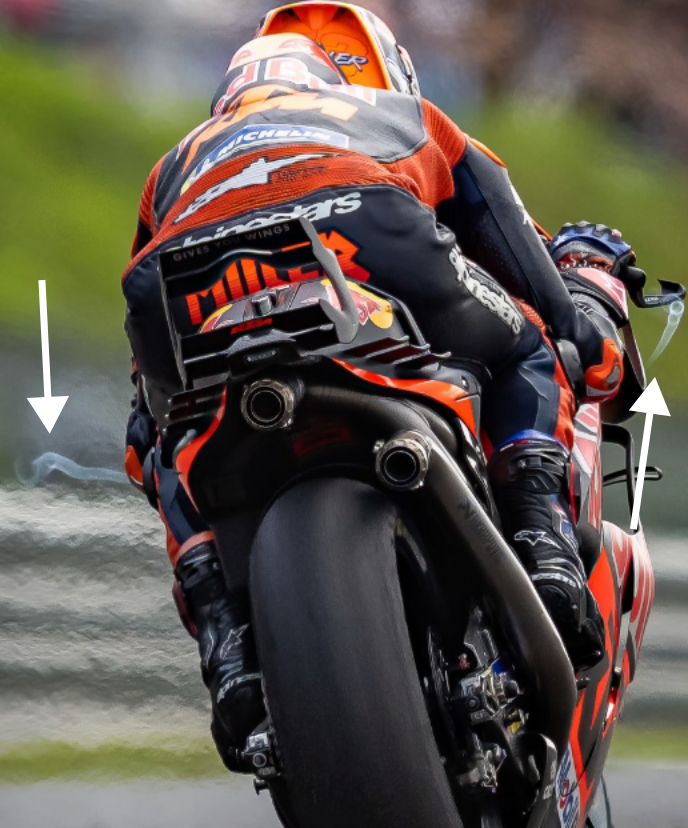}
\caption{Wingtip vortexes due to downforce generation on the Jack Miller's KTM RC16 2024 MotoGP prototype \cite{ktm_wingtip_vortex}.}
\label{fig_19}
\end{figure}

\begin{figure}[ht!]
\centering
\includegraphics[width=\columnwidth]{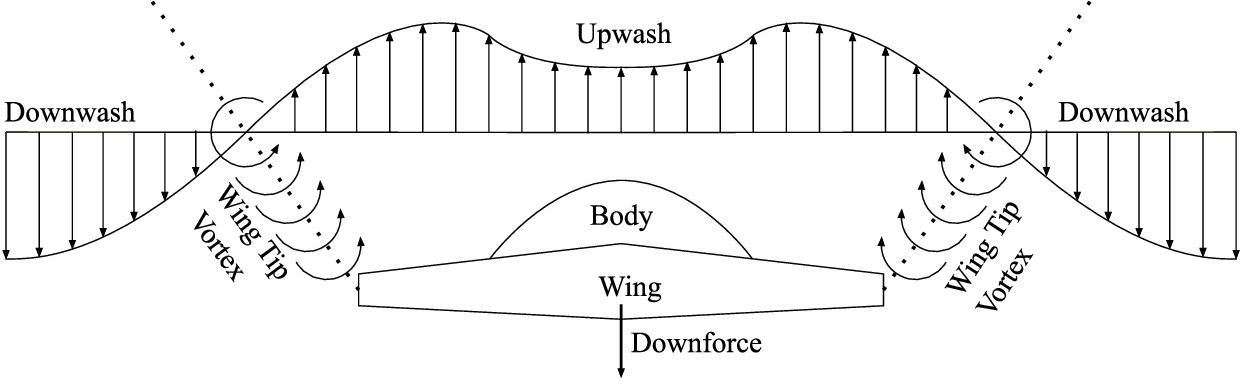}
\caption{Induced vertical velocity schematic produced by a wing generating downforce with a body at the wing center.}
\label{fig_04}
\end{figure}

The wing tip vortexes influence all the surrounding flow convected downstream, as can be seen in figure \ref{fig_19}. Wings fitted on the fairing's front section generates secondary flow components perpendicular to the main flow direction. According to figure \ref{fig_04}, for a downforce generating device, an upwash component will be generated and convected over the body and wing-aligned direction, while the outside section will be affected by a downwash convected component. The body presence usually reduces the lift distribution over the wing, reducing its overall lift \cite{raymer_2018}. The induced drag is proportional to the squared value of the generated lift while the upwash/downwash component follows the same proportion as the lift force \cite{anderson_2017}. The wingtip vortexes form a coherent structure that produces an induced angle of attack effect over the pursuing motorcycle. This may affect its performance positively or negatively, depending on its lateral relative position. The wash component may also induce less or more downforce production by the pursuing motorcycle wings and affect its naturally generated lift.


The aerodynamic coefficients analyzed in this work are defined as:
\begin{gather}
    C_L = \frac{L}{\frac{1}{2} \rho_\text{ref} U_\infty^2 A_\text{ref}}, \quad C_D = \frac{D}{\frac{1}{2} \rho_\text{ref} U_\infty^2 A_\text{ref}}, \nonumber \\
    C_M = \frac{M}{\frac{1}{2} \rho_\text{ref} U_\infty^2 A_\text{ref} L_{\text{ref}}}, \nonumber
\end{gather}
where the forces and their respective coefficients are the lifting force $L$ and the drag force $D$. The moment analyzed in this work is the pitching moment $m$. The reference values are the air density $\rho_{\infty}$, taken as incompressible in the numerical simulations, the undisturbed flow velocity $U_\infty$, the reference frontal area $A_{\text{ref}}$ and the reference wheelbase length $L_{\text{ref}}$. The pitching moment center is located on the ground at the rear tire contact patch, vertically aligned with the wheel spinning axis.

\begin{figure}[ht!]
\centering
\includegraphics[width=\columnwidth]{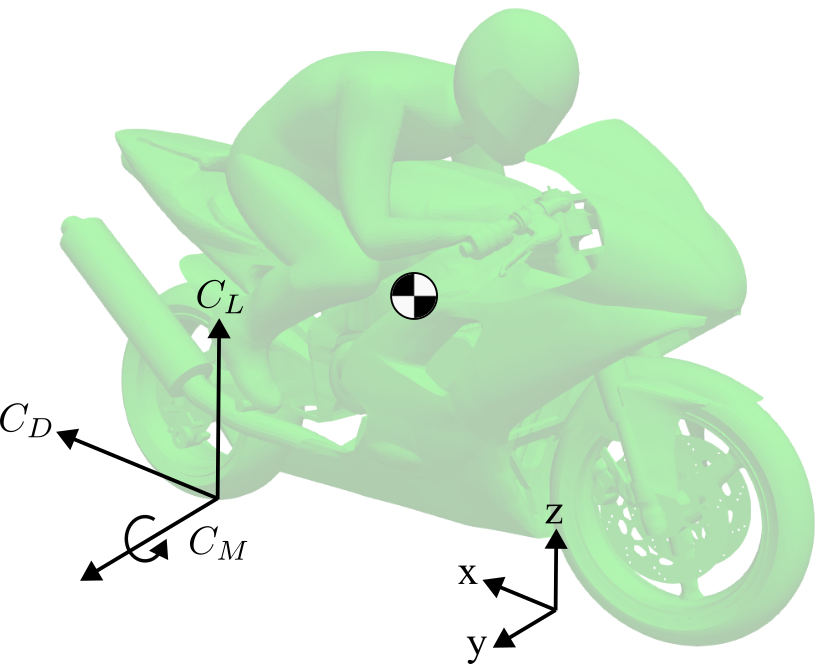}
\caption{Schematic representation of the aerodynamic generated forces and moments and their placement.}
\label{fig_05}
\end{figure}

\section{Numerical Setup}
\subsection{Mathematical Model and Numerical Method}

The governing equations used in the numerical simulations are the time filtered continuity and the three-dimensional form of the Navier-Stokes equations \cite{aristeu_2020,pope_2013} for the RANS treatment. The aerodynamic fluid flow is modeled after a Newtonian fluid with constant thermodynamic properties for the stress tensor. Since the obtained speed in the flow field solutions is far below the speed of sound for typical air conditions ($\approx 347$ m/s) \cite{anderson_2017}, the fluid model used in the numerical simulations is incompressible. Heat conduction effects were also not considered since they are typically negligible in low Mach flows. The resulting governing equations are given by
\begin{equation}
    \nabla \cdot \vec{u}_0 = 0 \quad \text{and}
    \label{eq_01}
\end{equation}
\begin{multline}
    \frac{\partial \vec{u}_0}{\partial t} + \nabla \cdot \left[ \vec{u}_0 \vec{u}_0 \right] = - \frac{1}{\rho_\text{ref}} \nabla p_0 + \\
    \nabla \cdot \left[ \nu_\text{ref} \left( \nabla \vec{u}_0 + \nabla \vec{u}_0^T \right) + \vec{\vec{\tau}}_{t} \right],
    \label{eq_02}
\end{multline}
where $\vec{u}_0$ is the mean velocity vector, $p_0$ is the mean pressure, $\rho_\text{ref}$ is the constant fluid density, $\nu_\text{ref}$ is the fluid constant kinematic viscosity and $\vec{\vec{\tau}}_{t}$ is the Boussinesq-Reynolds tensor, given by
\begin{equation}
    \vec{\vec{\tau}}_{t} = \nu_t \left( \nabla \vec{u}_0 + \nabla \vec{u}_0^T \right) - \frac{2}{3} k \vec{\vec{I}},
\end{equation}
where $k$ is the specific turbulent kinetic energy \cite{aristeu_2020} and $\nu_t$ is the turbulent kinematic viscosity, arising from Boussinesq's hypothesis. Since the terms refer to the problem of closing the average equations, their modeling is necessary for a dynamic effect on the associated transport phenomena. The model used for turbulent viscosity and kinetic energy in this work is the k-$\omega$ Shear Stress Transport (k-$\omega$-SST) model. The model constants used are the standard ones indicated in the original model work \cite{sst_94}. The governing equations \ref{eq_01} and \ref{eq_02}, along with the turbulence model are then integrated into control volumes to undergo numerical treatment and computational implementation.

The numerical method used in the simulations is the Finite Volume Method, with its implementation in the OpenFOAM CFD code version v2112. The numerical scheme is as follows: steady state of the governing equations, where all partial time derivatives are set to zero and only the integral spatial terms are solved. The integrated divergence terms related to the turbulence model are obtained by the \emph{upwind} scheme, the nonlinear convection term is calculated by the \emph{limitedLinearV} scheme and the Laplacian terms are calculated via the \emph{linear} interpolation scheme. The scheme used to solve the resulting integral form of the incompressible form of the governing equations is the SIMPLE algorithm \cite{caretto_1972}. 

The simulations were run in the Amadea HPC (High Performance Computing) cluster from the Aeroacoustics Laboratory (CaaLab) of the University of Brasília. The Amadea is a Beowulf computer cluster with eight compute nodes used in the calculations. Each node has an Intel Xeon Phi 7210 with 64 cores and 256 threads, each equipped with a total of 116 GB of RAM memory. They are interconnected to a manager and storage node in an InfiniBand FDR-type network. The cluster is managed and scheduled through the PBS (Portable Batch System) under the Centos Linux 7.2 operating system. The OpenFOAM code was compiled on the GCC compiler version 11.2 and the MPI library used for interprocess communication is the MVAPICH2 version 2.3.7 \cite{mvapich2}. Several decompositions were tested for minimum computing time and the tested number of MPI processes were 32, 50, 64, 100, 128, 150, 200 and 256. The fastest decomposition was of 128 MPI processes for distributed computing and each iteration took approximately four seconds.

\begin{figure}[ht!]
\centering
\includegraphics[width=1.0\columnwidth]{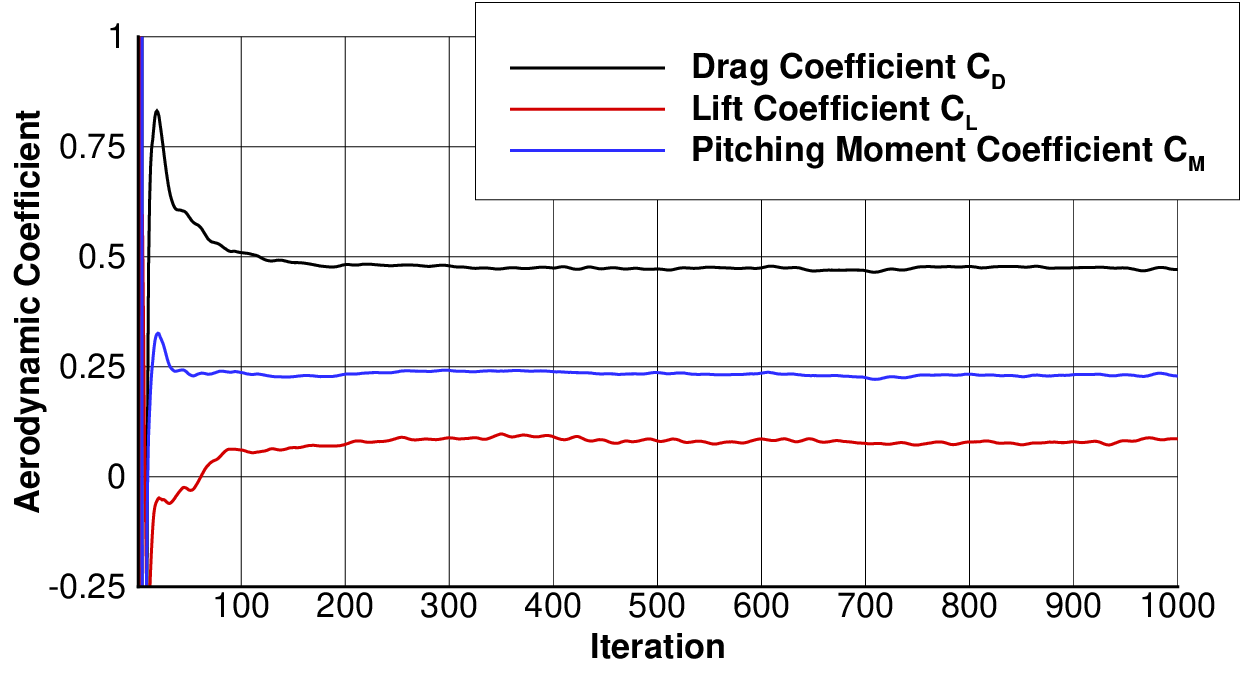}
\caption{Coefficient convergence for the three aerodynamic coefficients for the wingless single motorcycle case.}
\label{fig_17}
\end{figure}

In the single motorcycle case, 1000 iterations were run to achieve convergence, while in the simulations where the aerodynamic influence on the trailing motorcycle is assessed, at least 2000 iterations were run to achieve convergence. The force and moment coefficients of the first half of iterations were discarded and the mean coefficients values windows are taken from the last half of iterations. The aerodynamic coefficient as a function of numerical iteration for the single wingless motorcycle case is given in figure \ref{fig_17}. Convergence can be observed early on the simulation and the same trend is obtained in all the numerical simulations of this work. Further case configuration and computational environment details are available upon request.

\subsection{Computational Discretization and Flow Conditions}

The simulated motorcycle model is the Yamaha YZF-R1, a production sports bike of the superbike class (1000cc), released in 1998. At the time of its launch in 1998, superseding the YZF1000R Thunderace, the YZF-R1 presented technologies that were race-aimed: inverted suspension forks, a short wheelbase for rapid transitions, aluminum chassis and an updated four-cylinder four-stroke liquid-cooled engine with a maximum power of $148.8$ hp, achieving a claimed top speed of $270$ km/h and with a dry weight of 177 kilograms. It followed the racing counterparts in performance and aerodynamic trends, being equipped with a full fairing and with racing body position for the pilot.

\begin{figure}[ht!]
\centering
\includegraphics[width=1.0\columnwidth]{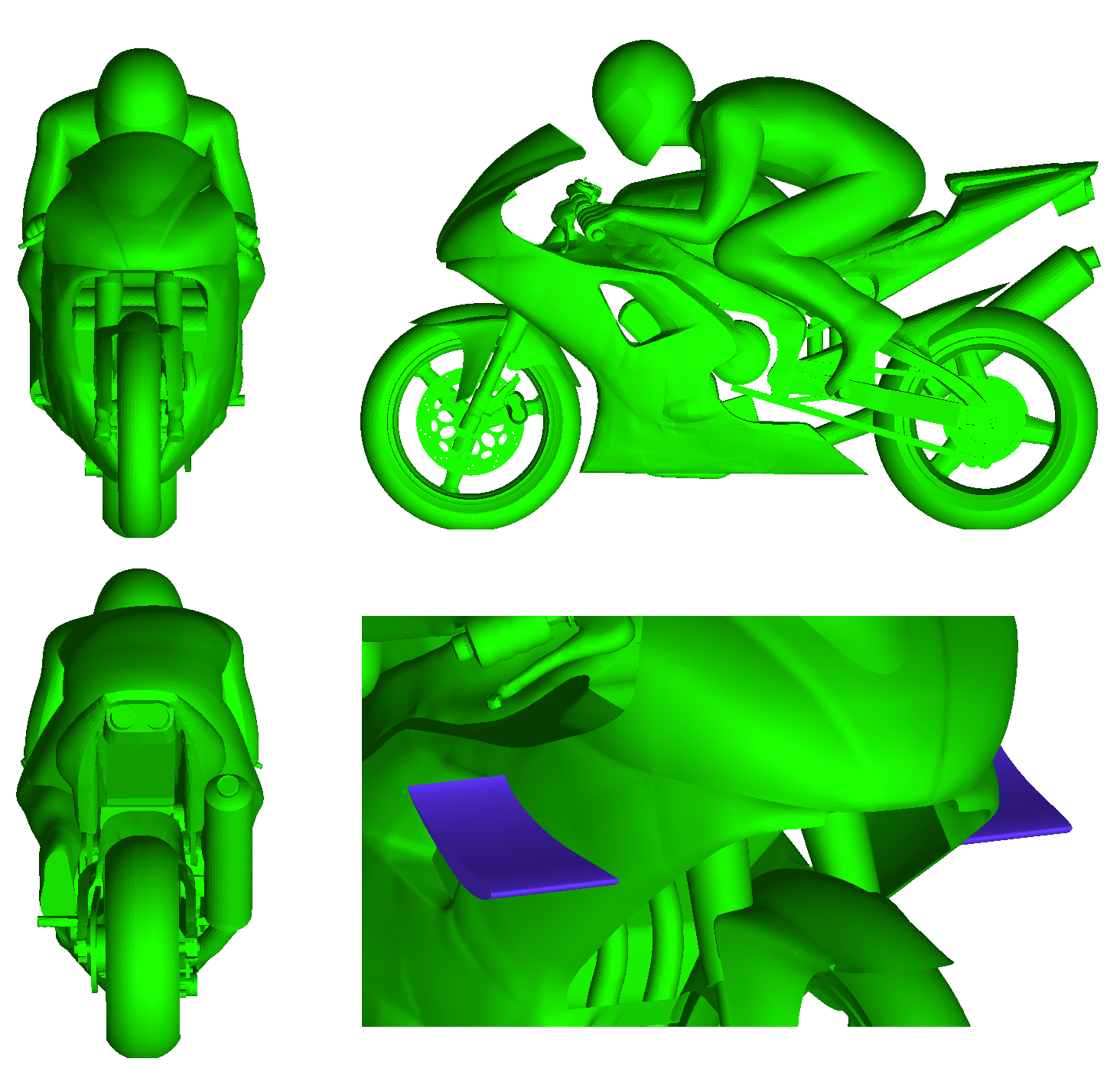}
\caption{CAD model of the Yamaha YZF-R1 used in the simulations and the wing mounted (blue) fairing for aerodynamic influence simulations.}
\label{fig_06}
\end{figure}

\begin{figure*}[ht]
\centering
\includegraphics[width=1.0\columnwidth]{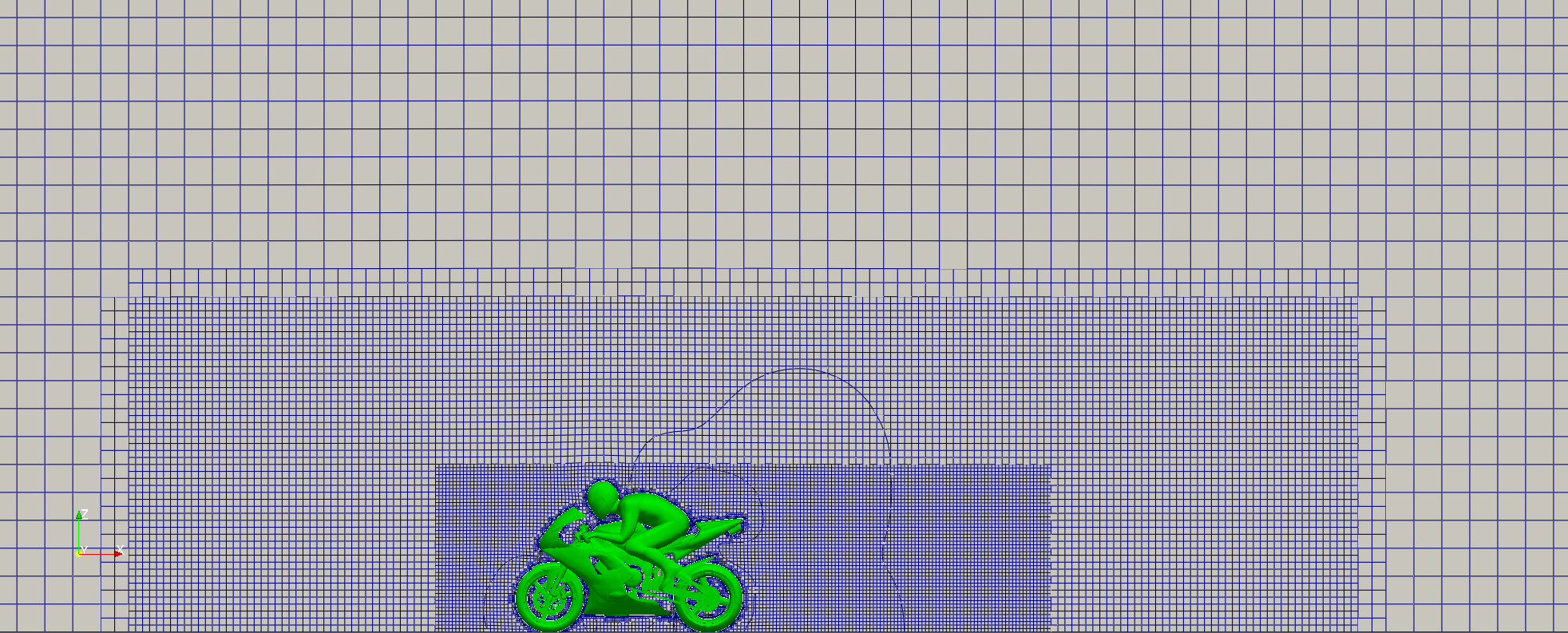}
\includegraphics[width=1.0\columnwidth]{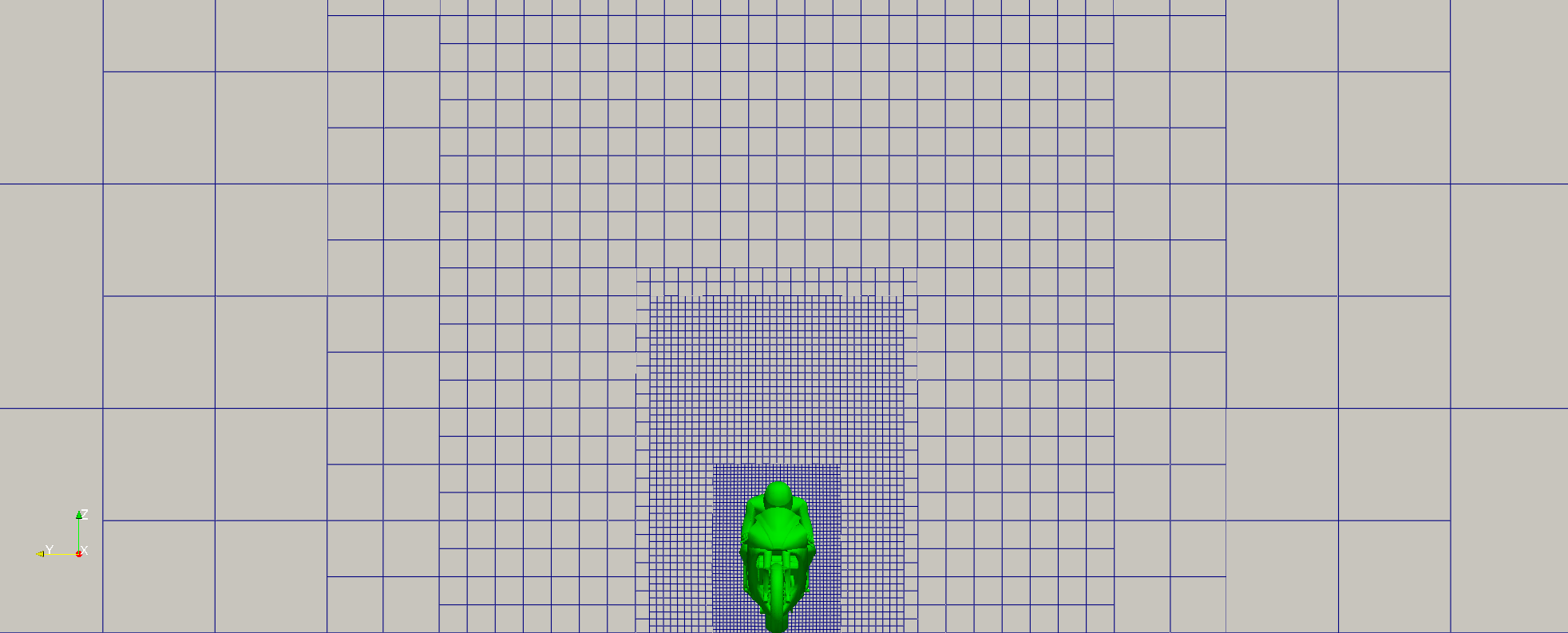} \\
\includegraphics[width=0.969\linewidth]{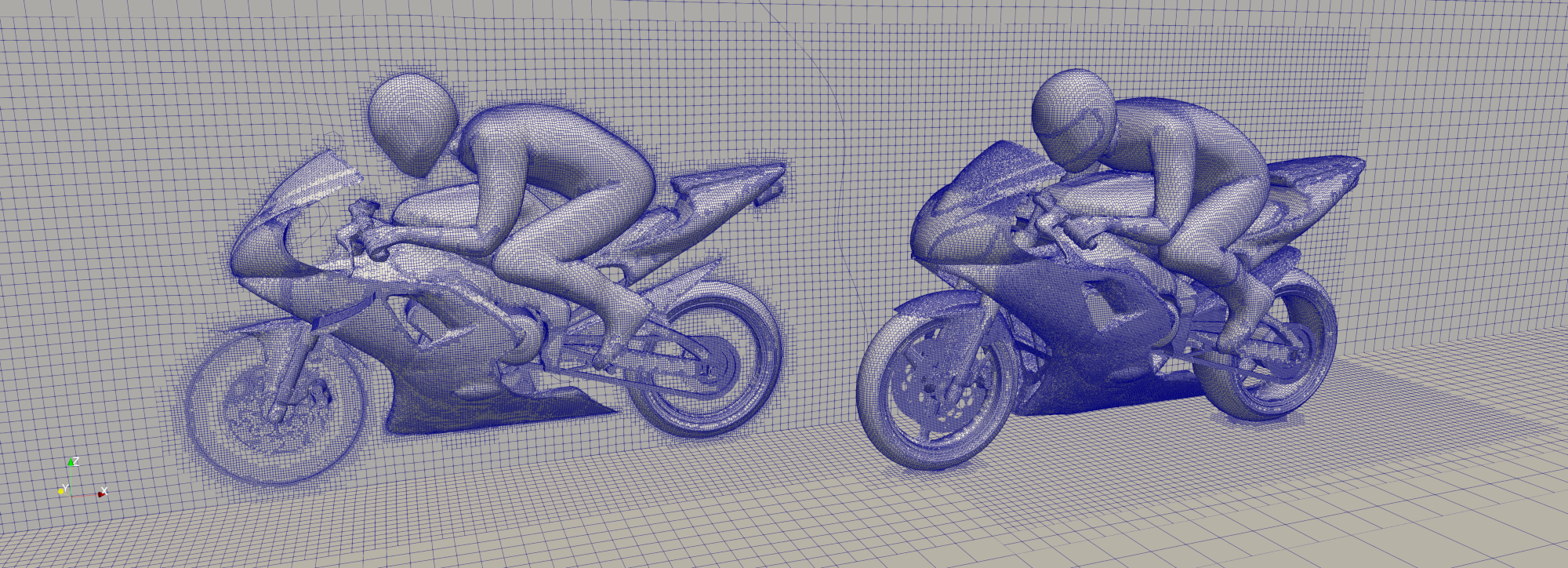}
\caption{Lateral and front plots of the generated simulation mesh showing the refinement regions with one motorcycle only mesh (upper) and the nearfield surface mesh for the two motorcycles case with lateral position displacement (lower).}
\label{fig_07}
\end{figure*}

The CAD model of the YZF-R1 is easily obtained from the OpenFOAM tutorials, where a low-resolution LES case can be run for code testing. The CAD model resembles a racing specification motorcycle, where its mirrors, license plate and turning signals were removed and a pilot in tucked position was added. The typical values for geometry are similar to MotoGP prototypes and race-spec WSBK modern motorcycles. Figure \ref{fig_06} shows the CAD model of the YZF-R1 used in the simulations along with the added wing in the front fairing. For the wing-fitted motorcycle simulations, a simple one-body profile wing was added to the CAD model. The wing has a constant profile Selig-1223 aerodynamic airfoil with constant spanwise angle of attack of -40$^\circ$ in respect to ground movement and a constant chord of 0.2 m. The semi-wing has a span of 0.1636 m. The motorcycles are located at the center of the numerical mesh relative to the $x$ and $y$ axes and at the boundary of the lower $z$ axis boundary, as depicted in figure \ref{fig_07}.

\begin{figure}[ht!]
\centering
\includegraphics[width=0.9\columnwidth]{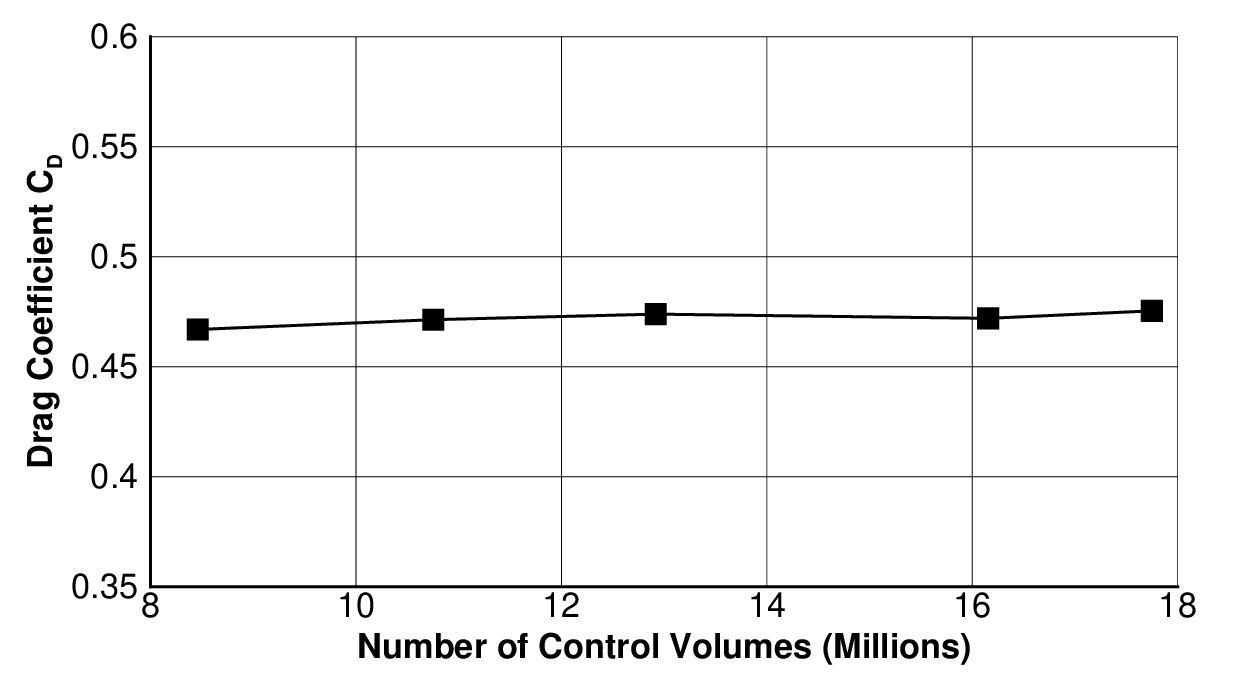}
\caption{Mesh convergence study of the drag coefficient for a single motorcycle simulation.}
\label{fig_08}
\end{figure}

Figure \ref{fig_08} shows the drag convergence study to define the mesh generation parameters. Strong convergence may be observed since the coarsest mesh with $\approx 8$ millions control volumes, follows the same trend in numerical testes with $\approx 18$ million volumes. Meshes targeted at around 12 million volumes were chosen to be used in the single motorcycle simulations, showing a compromise between space resolution and simulation cost. The dual motorcycle meshes resulted in meshes ranging from 21 million up to 25 million control volumes.

The numerical meshes used in the simulations were generated by the \emph{blockMesh} and the \emph{snappyHexMesh} utilities provided by the OpenFOAM package. Since mesh generation in the \emph{snappyHexMesh} is made in a top-down approach, \emph{blockMesh} resulting block is the mesh input. A single space structured block is provided with the coordinate system shown in figure \ref{fig_05} and the following dimensions and volume resolution:
\begin{itemize}
    \item 300m in the streamwise $x$-direction.
    \item 120m in the lateral $y$-direction.
    \item 80m for the total height in the $z$-direction.
    \item far-field regular volume distribution with 4 m sided cube shaped volumes.
\end{itemize}

The turbulent wake profile generated by the leading motorcycle persists over some characteristic lengths after complete dissipation \cite{blocken_2020,winski_2022} and must be modeled accordingly. Around the region of interest containing the motorcycle geometry, a refinement box was added in order to accommodate smaller control volumes. Its refinement level is 4, i.e. the volumes are $4/2^4 = 0.25$ m in length. It is located at the center of the \emph{blockMesh} generated box and at the ground level. Its dimensions are $21.5$ m in the $x$-direction, 6.3 m in the $y$-direction and 5.2 m from the ground upwards in the $z$-direction. Another refinement box was added inside the former with dimensions of 11 m by 2.2 m by 2.6 m. Its refinement level is 6, resulting in control volumes with $4/2^6 = 0.0625$ m in length. Each motorcycle will then be contained in its own refinement box with its center located at the aft of the rear tire. Its dimensions are 5.5 m by 1.1 m by 1.3 m, with a refinement level of $7$, resulting on control volumes with $4/2^7 = 0.03125$ m in length. Values of longitudinal distance from $x=5$ cm and up to $7$ m will be evaluated since good agreement can be observed for RANS simulations when compared to more accurate simulations, such as the RANS-LES hybrid methods \cite{blocken_2020,winski_2022}. The lateral distances will range from $y=0$ cm up to 60 cm with a fixed longitudinal relative distance of $x=5$ cm.

The motorcycles boundary layer mesh has 5 volumes in the normal direction, providing a theoretical $y^+$ value based on flat plate turbulent boundary layer of $y^+=10$ at $x=1.42$. The majority of drag composition of the turbulent wake is due to profile drag thus no further reduction of $y^+$ was necessary, as it did not affect the overall drag in the showed mesh convergence study.

The reference numerical values used in the numerical simulations are given in table \ref{tab_01}. The resulting Reynolds number of the numerical simulations is $Re = U_\infty L_\text{ref}/\nu_{\text{ref}} = 6.992\cdot10^{6}$.

\begin{table}[ht!]
\begin{tabular}{c|c}
Reference Physical Quantity & Numerical Value    \\
\hline
$\rho_{\text{ref}}$ (kg/m$^3$)  & $1.2886$                  \\
$U_{\infty}$ (m/s)              & $70.0$                    \\
$A_{\text{ref}}$ (m$^2$)        & $0.644384$                \\
$L_{\text{ref}}$ (m)            & $1.42$                    \\
$p_{\infty}$ (Pa) (relative pressure)               & $0.0$  \\
$\nu_{\text{ref}}$ (m$^2$/s)    & $1.436\cdot10^{-5}$       \\
Front Wheel Angular Velocity    & $-225.8064$ rad/s         \\
Rear Wheel Angular Velocity     & $-229.5081$ rad/s         \\
\end{tabular}
\caption{Geometric and thermodynamic values of the reference quantities.}
\label{tab_01}
\end{table}

Even though the simulated cases are steady state, the iterative numerical scheme needs a set of initial conditions in the convergence process. Uniform fields for the static pressure of $p=0.0$ Pa and flowfield freestream velocity of $\vec{U} = 70.0 \hat{\imath} + 0.0 \hat{\jmath} + 0.0 \hat{k}$ are imposed as initial conditions. The used levels of turbulent viscosity $\nu_t$, turbulent kinetic energy $k$ and the turbulent specific dissipation rate $\omega$ are the typical ones for the k-$\omega$-SST RANS model \cite{k-omega-sst_openfoam}.

The boundary regions applied on the far-field block generated by the \emph{blockMesh} utility are grouped into inlet, outlet, lateral, upper and ground regions. Normal zero gradient of static pressure pressure is imposed at the lateral, upper, ground and inlet regions, while at the outlet the static pressure fixed value of $0.0$ Pa is imposed. The velocity boundary conditions are full slip for the lateral and upper regions and fixed freestream values at the inlet and ground. Zero gradient velocity is applied on the outlet region for outflows and the no backflow constraint is also imposed when the velocity field reaches negative values for the $u \hat{\imath}$ component. The turbulence values through the inlet are the same as the initial condition values.

Along the motorcycles surface, the no-slip boundary condition is imposed for the velocity, while at its rotating regions, wheels, tires and brake discs, the surface velocity is derived from the their respective angular velocities listed in table \ref{tab_01}. A normal zero pressure gradient is imposed over all the motorcycle surfaces. Since the $y^+$ obtained in the boundary layer mesh is above the recommended values for typical RANS values, the wall function is used for the $k$ \cite{k_wall_function}, $\omega$ \cite{omega_wall_function} and $\nu_t$ \cite{nu_t_wall_function} parameters.

\section{Results}

As a baseline for coefficient comparison, a wingless and wing-equipped single motorcycle cases were simulated to establish base values. These base aerodynamic coefficients are given in table \ref{tab_02}. The drag, lift and pitching moment are decomposed into the pressure and shear stresses in both cases. As expected, most of the aerodynamic forces result from pressure distribution, due to flow separation in the motorcycle and rider aft region, in concordance with previous works \cite{araki_2001, takahashi_2009, watanabe_2003, fintelman_2015}. The wing-equipped motorcycle produces more drag due to the downforce generation and the resulting induced drag, also resulting in an intended sign change in lift and reduction of the pitching moment.

\begin{table}[!htp]
\centering
\begin{tabular}{c|c|c|c}
Wingless case      & $C_D$ & $C_L$ & $C_M$           \\
\hline
Pressure Stress    & 0.45733 & 0.07999  & 0.22298    \\
Viscous Stress     & 0.01661 & -0.00071 & 0.00801    \\
Total stress       & 0.47394 & 0.07928  & 0.23099    \\
\hline
Wing-equipped case & $C_D$ & $C_L$ & $C_M$           \\
\hline
Pressure Stress    & 0.51329 & -0.16609 & 0.07873    \\
Viscous Stress     & 0.01727 &  0.00064 & 0.00919    \\
Total stress       & 0.53056 & -0.16545 & 0.08791
\end{tabular}
\caption{Baseline aerodynamic coefficients of a single motorcycle without and with wings according to the type of stress and the overall value.}
\label{tab_02}
\end{table}

\begin{figure}[ht]
\centering
\includegraphics[width=0.99\columnwidth]{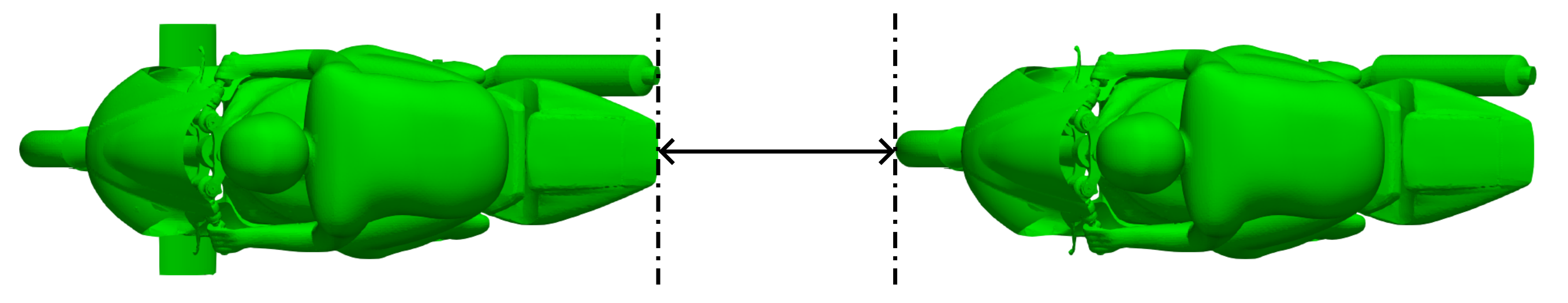} \\
\vspace{7mm}
\includegraphics[width=0.99\columnwidth]{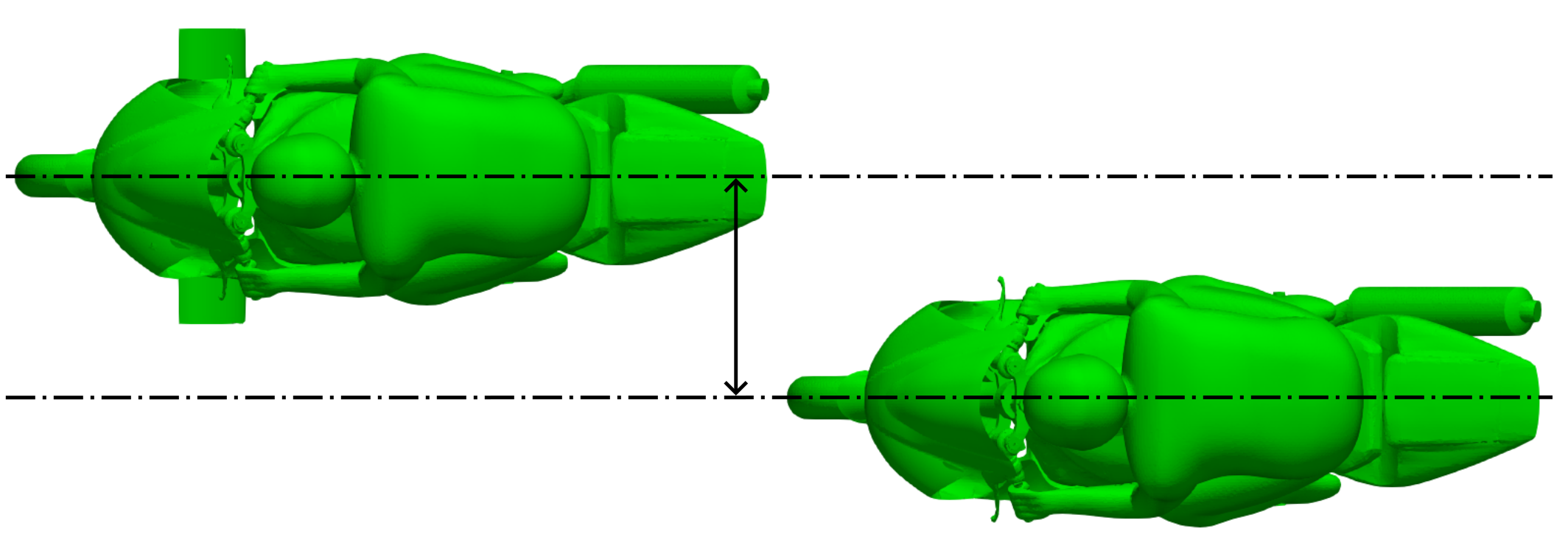}
\caption{Longitudinal (upper figure) and lateral (lower figure) displacement of relative position of the leading and the trailing motorcycles.}
\label{fig_18}
\end{figure}

The aerodynamic influence assessment between two trailing motorcycles will be presented in a systematic way, where the longitudinal and lateral distances (figure \ref{fig_18}) influence over the aerodynamic coefficients will be evaluated along with three main cases:
\begin{enumerate}
    \item Case 1: leading and trailing motorcycles without wings.
    \item Case 2: front motorcycle with and without wings and the trailing motorcycle without wings.
    \item Case 3: both motorcycles equipped with wings.
\end{enumerate}

\subsection{Case 1: Simulations With Both Wingless Motorcycles}

The first case assessment is both motorcycles with no wings and the influence of the longitudinal distance on the dynamic coefficients. The longitudinal distance between the motorcycles is the measured $x$-axis distance between the rear tire of the leading motorcycle and the front tire of the pursuing one with no lateral deviation.

In a hypothetical competitive environment case of the overtaking process on a straight line, the leading motorcycle is approached from the rear by a drafting motorcycle where these are almost aligned longitudinally and then a lateral movement is taken to proceed with the overtake. In this case, the trailing motorcycle would be exposed to all the simulated relative positions and the presented aerodynamic coefficients should represent the flow regimes that it is subjected to. A similar race condition is the packing, where all motorcycles run along a single line, where the longitudinal distance variation dictates influence over the aerodynamic coefficients.

\begin{figure}[ht]
\centering
\includegraphics[width=0.45\columnwidth]{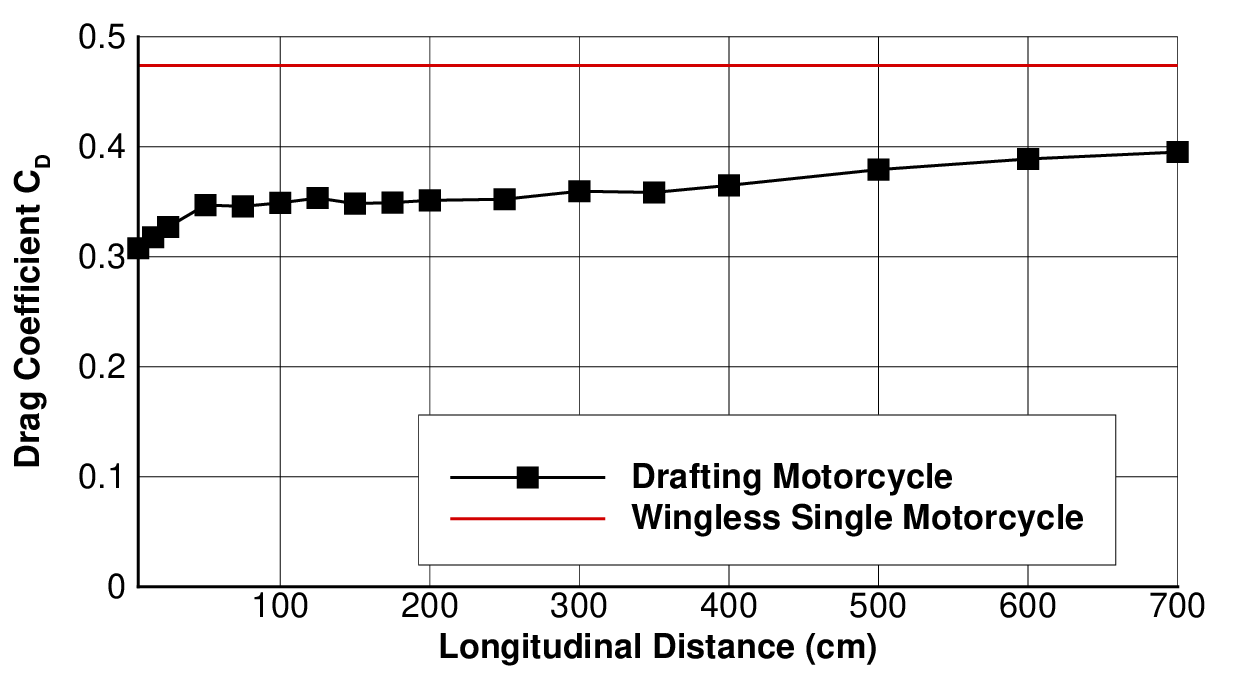}
\includegraphics[width=0.45\columnwidth]{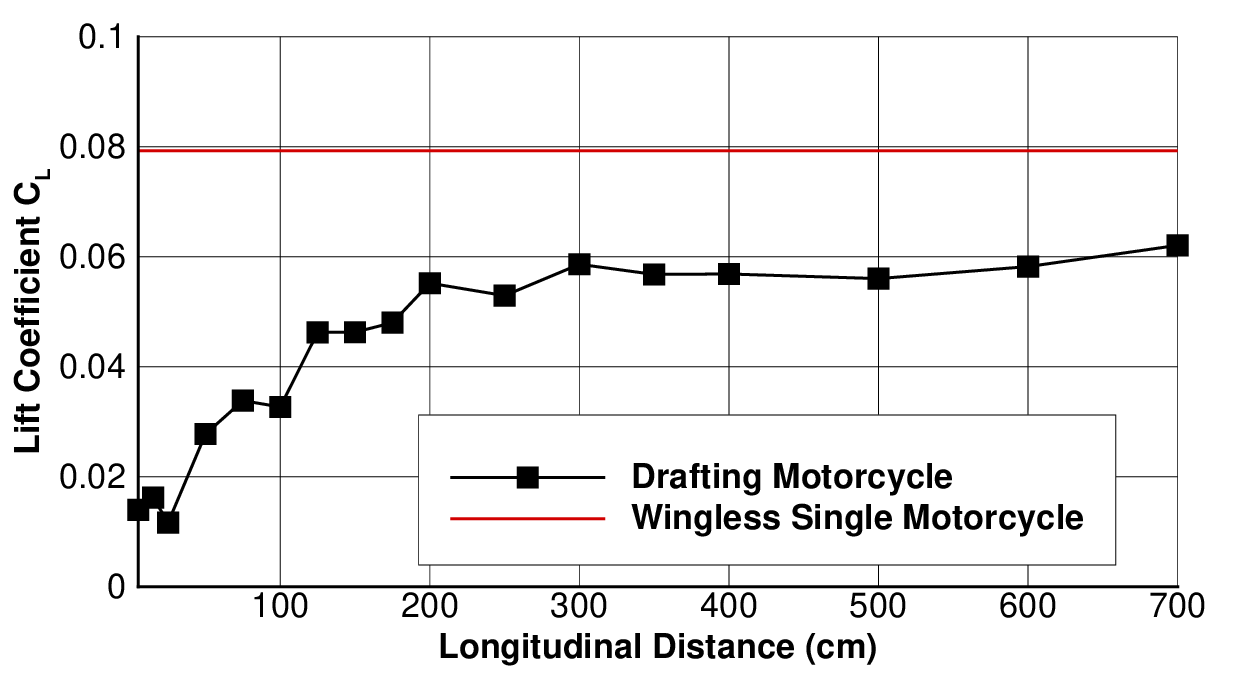} \\
\includegraphics[width=0.45\columnwidth]{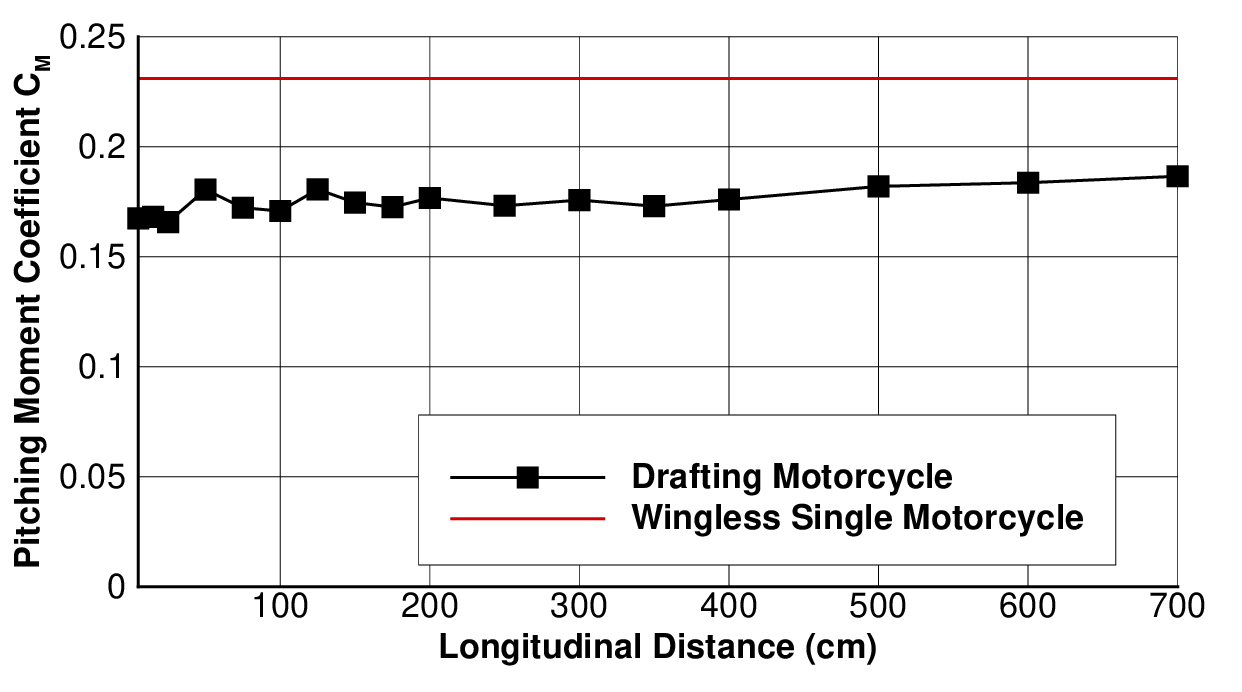}
\caption{Aerodynamic coefficients variation due to longitudinal distance between the leading and the drafting motorcycle of case 1.}
\label{fig_10}
\end{figure}

An asymptotic drag increase can be observed for the drafting motorcycle in figure \ref{fig_10}, while at the largest value, the base drag coefficient was not yet obtained. The low-pressure flow recirculation can still influence the downstream region several characteristic lengths further on. This same trend can be observed for the pitching moment, with a small oscillation at a closer distance. The lift coefficient exhibits strong variation in a close longitudinal distance up until $2$ m, then the same asymptotic behavior takes place as the distance between the motorcycles grows larger.

The reduction of all the presented dynamic coefficients provides a competitive advantage to the trailing motorcycle when in racing conditions. The overall drag reduction will facilitate overtaking since higher acceleration and top speed may be obtained, while better stability and reduced wheeling tendency may also be obtained from lift and pitching moment reduction, respectively. The presented case will be used as a baseline for the following cases, where the drafting motorcycles will have their resulting dynamic coefficients analyzed upon the aerodynamic influence of the front motorcycle equipped with downforce-generating wings.

\begin{figure}[ht]
\centering
\includegraphics[width=0.45\columnwidth]{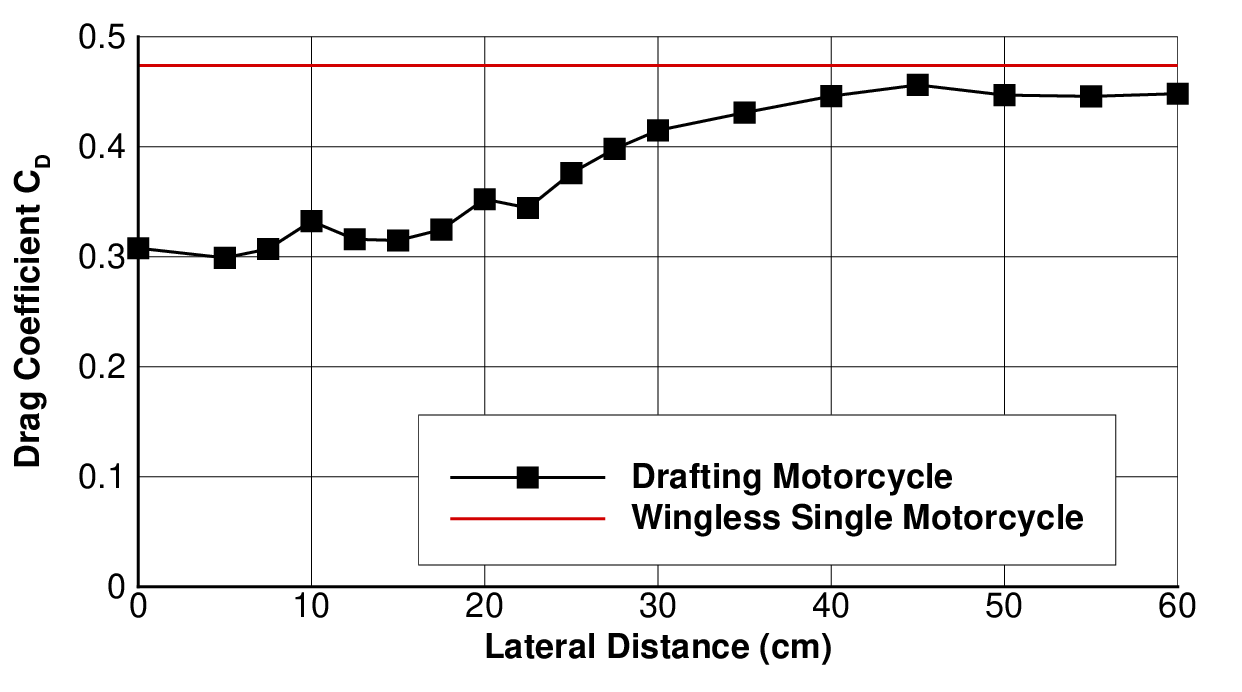}
\includegraphics[width=0.45\columnwidth]{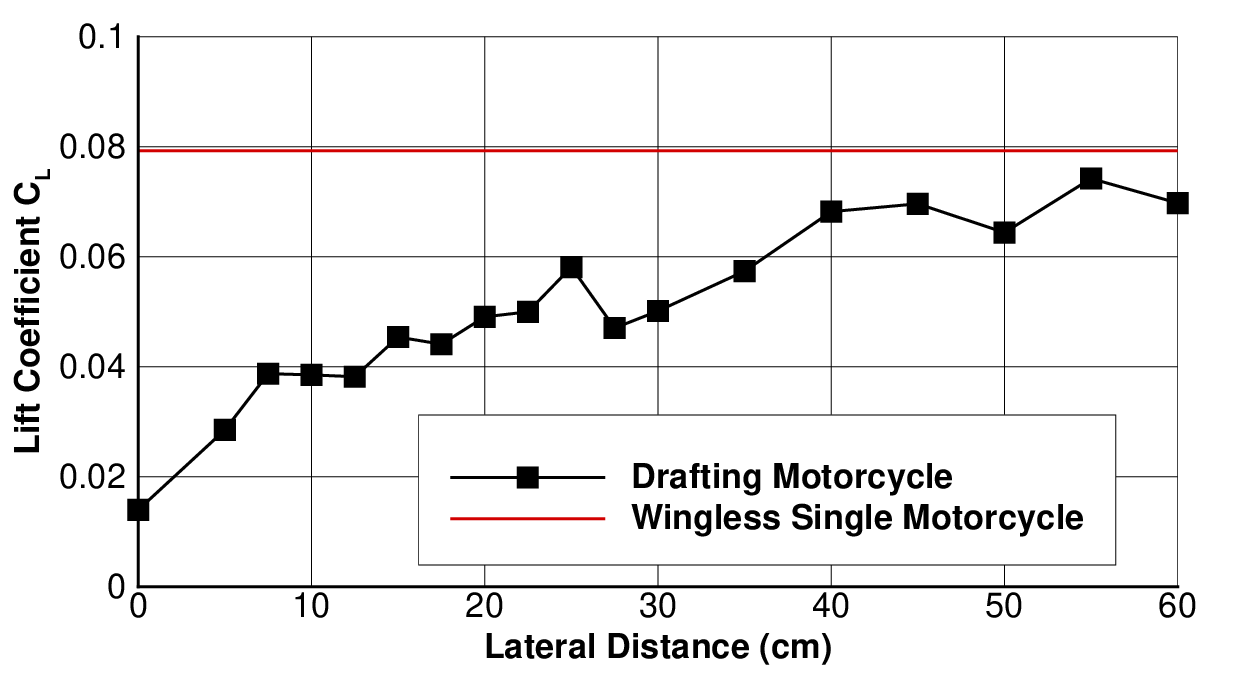} \\
\includegraphics[width=0.45\columnwidth]{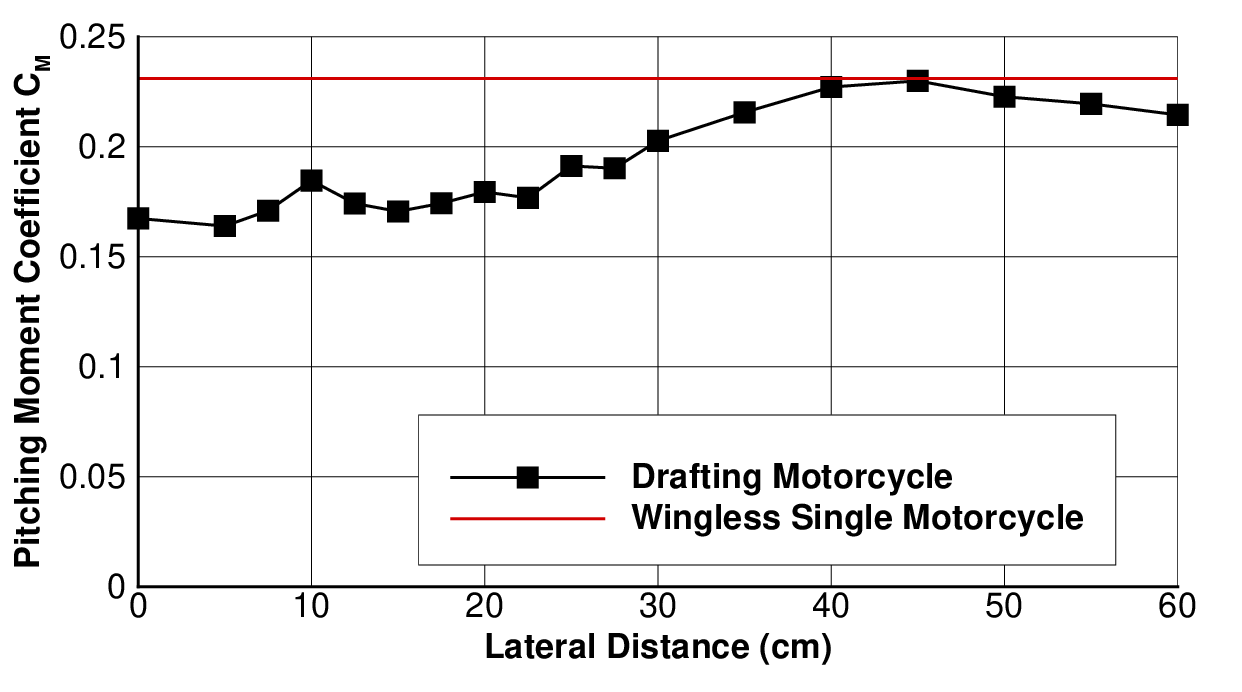}
\caption{Aerodynamic coefficients variation due to lateral distance between the leading and the drafting motorcycle of case 1.}
\label{fig_11}
\end{figure}

\begin{figure*}[ht]
\centering
\includegraphics[width=1.0\columnwidth]{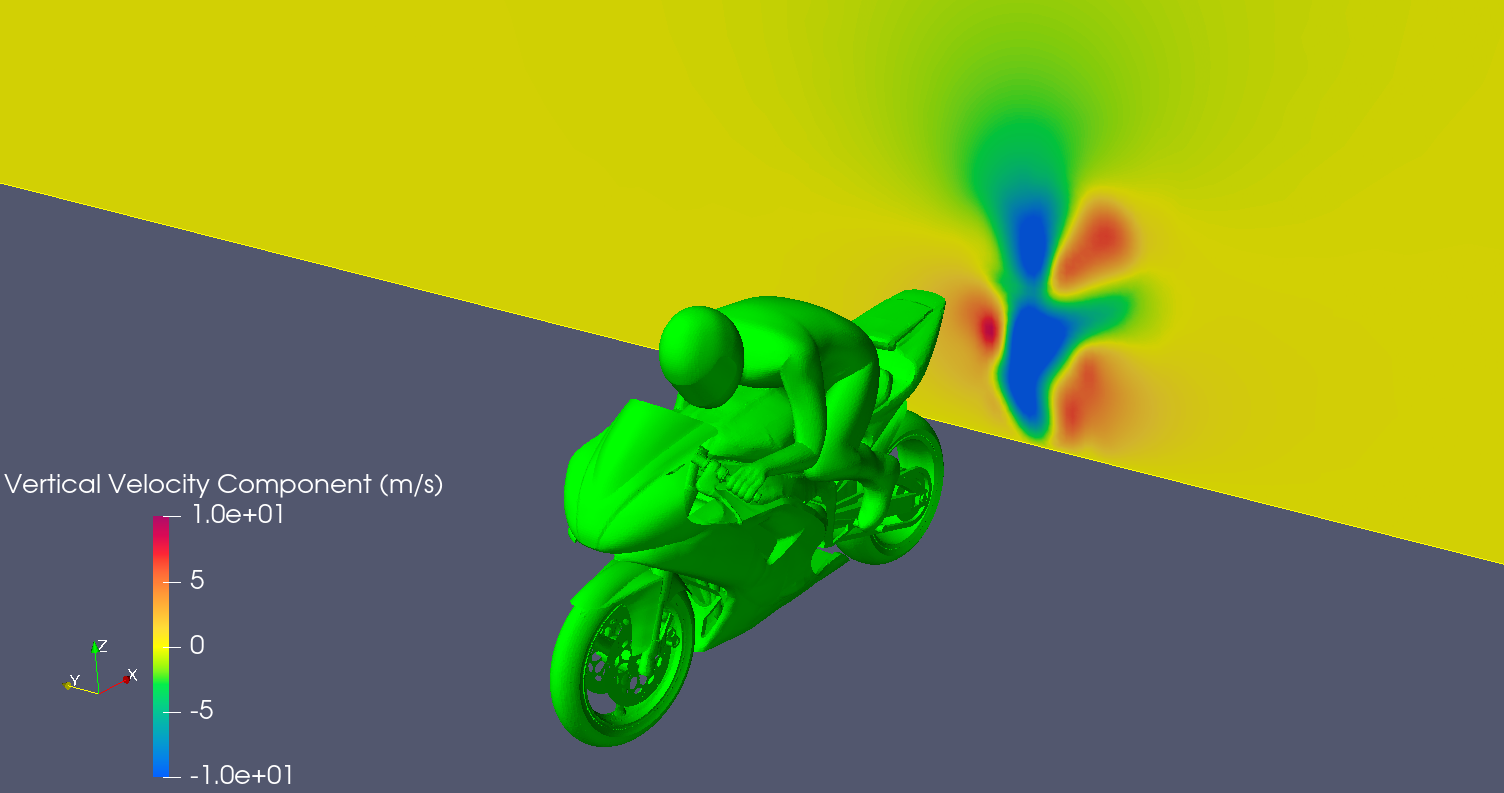}
\includegraphics[width=1.0\columnwidth]{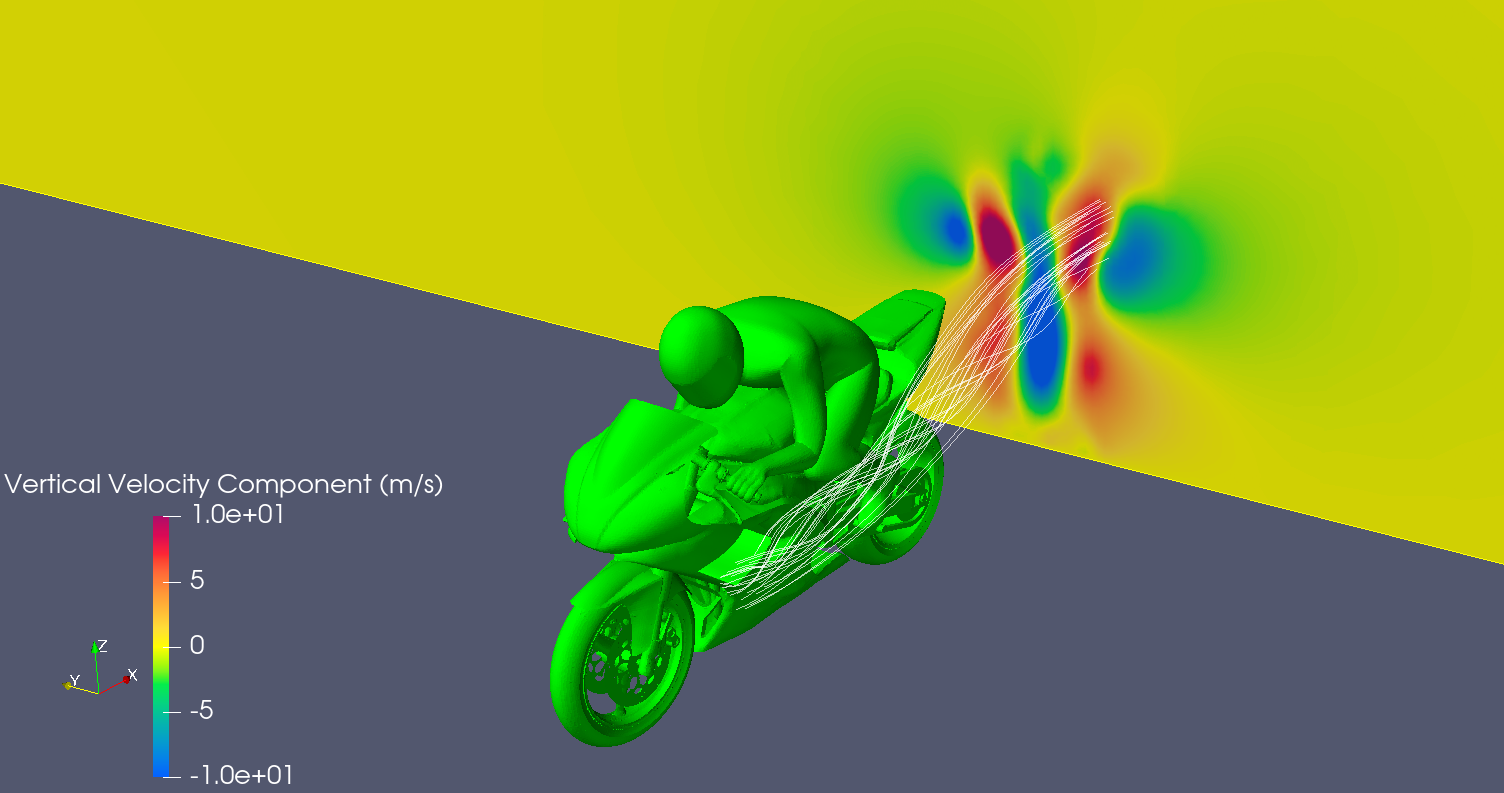}
\caption{Downstream flow visualization and vertical velocity component plotting on a plane perpendicular to the undisturbed flow direction for the wingless and wing-equipped configurations.}
\label{fig_14}
\end{figure*}

The dependency of the dynamic coefficients with respect to the lateral position is shown in figure \ref{fig_11}, ranging from totally aligned motorcycles up to $60$ cm of lateral deviation in the positive $y$ direction. The same trend of less aerodynamic influence as the relative distance grows can be observed for all the coefficients. In the pitching moment case, the single motorcycle coefficient is obtained at around $y=45$ cm but lessens for higher values of lateral distance. Some local oscillations can be observed for the drag and lift coefficients at closer distances, implying greater dynamic instability for the drafting motorcycle at the overtaking approach.



\subsection{Case 2: Aerodynamic Effect of the wing-Equipped Leading Motorcycle over the Wingless Pursuing Motorcycle}

This case aims to assess the aerodynamic influence of the wing-equipped leading motorcycle over the drafting one. Apart from the large turbulent wake region in the downstream of the leading motorcycle, coherent structures will arise from the downforce generation, resulting into a secondary flow on the flow direction perpendicular plane, as schematically indicated by figure \ref{fig_04}, seen by the vortex core low-pressure vapor condensation in figure \ref{fig_19} and visualized in a simulation in figure \ref{fig_14}. In the downstream direction of the wing-fitted leading motorcycle, an upward velocity component will be carried when laterally aligned, while for larger lateral displacement of the trailing motorcycle, a downwash velocity component will prevail as secondary flow on the plane perpendicular to the undisturbed flow. These coherent flow structures are mirrored in the $y$-direction, as expected for a symmetric lift-generating wing. In the wingless configuration, there is only the recirculating wake behind the motorcycle, resulting into a non symmetric perpendicular velocity field. These are the main flow features and their influence will be assessed in the aerodynamic coefficients of the following results.

\begin{figure}[ht]
\centering
\includegraphics[width=0.45\columnwidth]{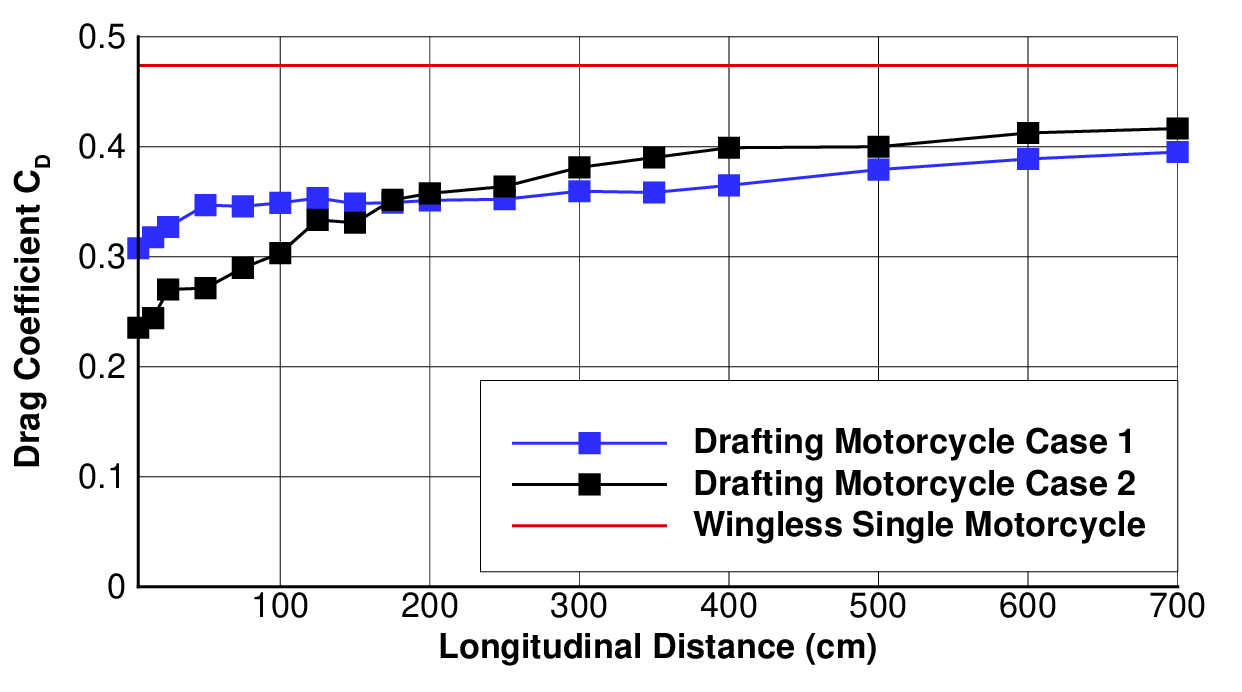}
\includegraphics[width=0.45\columnwidth]{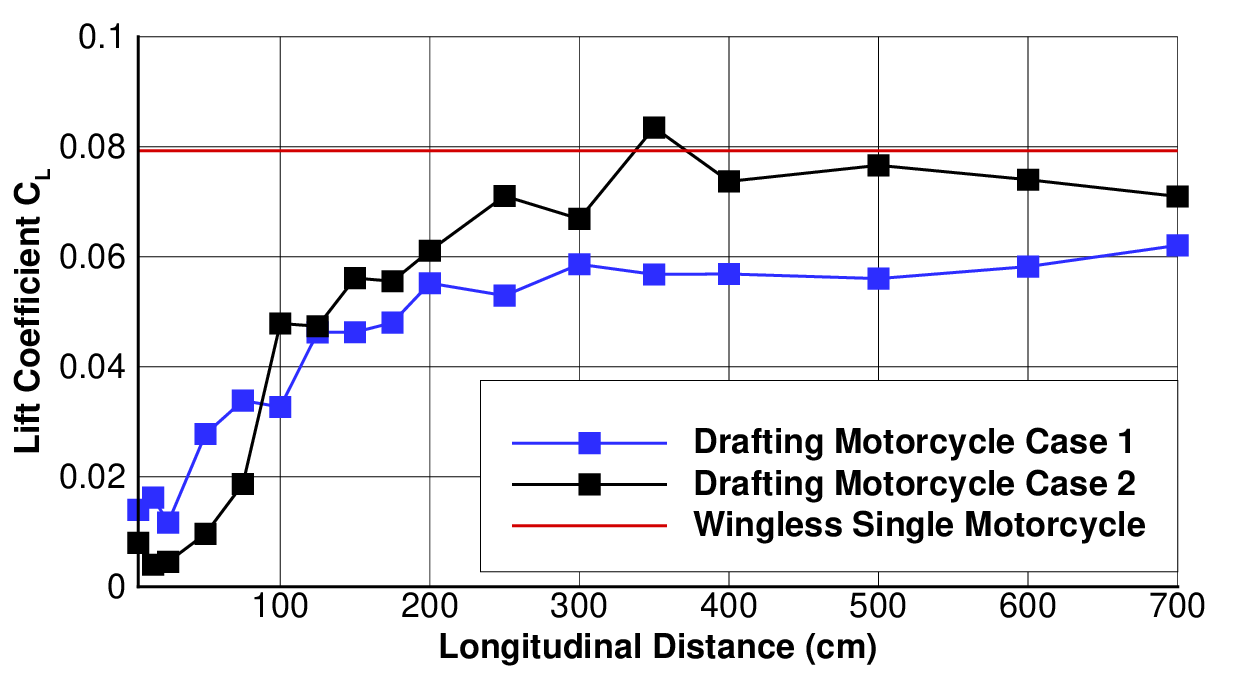} \\
\includegraphics[width=0.45\columnwidth]{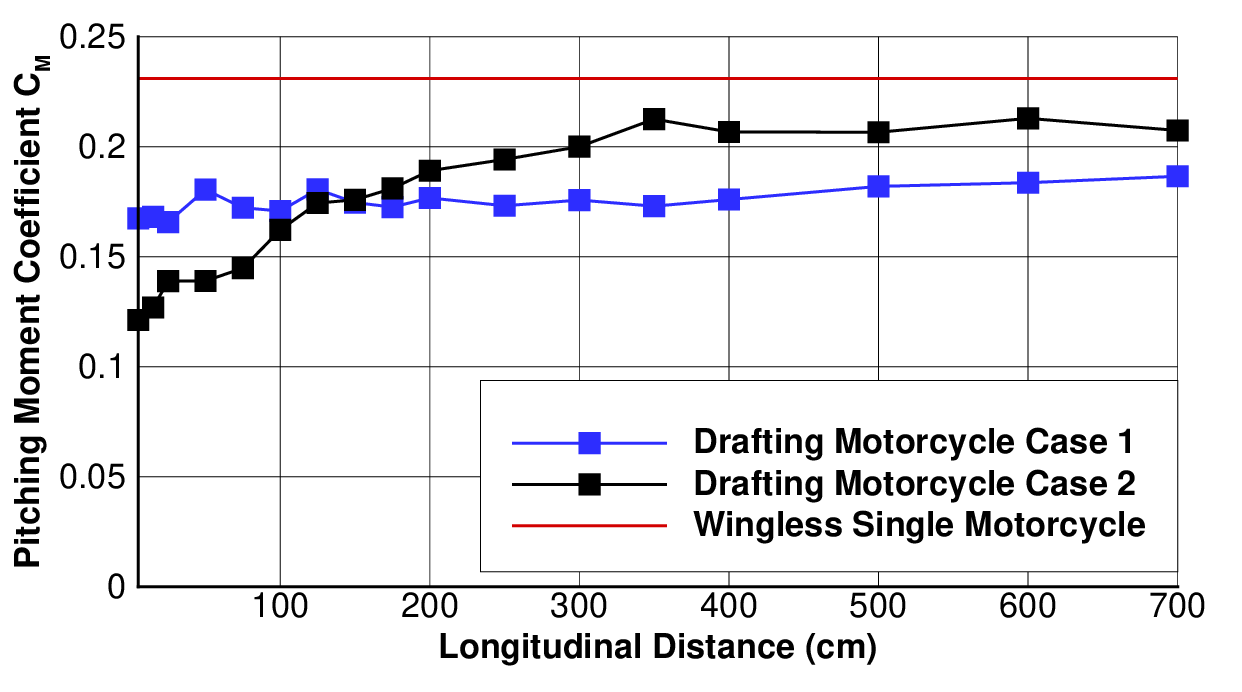}
\caption{Aerodynamic coefficients variation in function of longitudinal distance for the front motorcycle without wings (case 1) and with wings (case 2).}
\label{fig_12}
\end{figure}

Figure \ref{fig_12} shows The aerodynamic coefficients as a function of the longitudinal distance variation. At a closer distance from the leading motorcycle, the trailing motorcycle suffers a higher reduction for the drag, lift and pitching moment when compared to the flow regime in case 1, affecting its racing performance positively. But for larger longitudinal distances the drafting effect is somewhat smaller and the aerodynamic coefficients attain larger undesired values in case 2 when compared to case 1. This aggravates the stability and safety conditions, thus resulting into less racing advantage for the drafting motorcycle at the beginning of the approaching stage. This is due to the upwash velocity component generated by the wing-equipped leading motorcycle being convected and thus affecting the trailing motorcycle, resulting into an increase of the wheeling tendency. For larger longitudinal distances case 2 shows a less secure condition when compared to the aerodynamic influence of the wingless leading motorcycle.

\begin{figure}[ht]
\centering
\includegraphics[width=0.45\columnwidth]{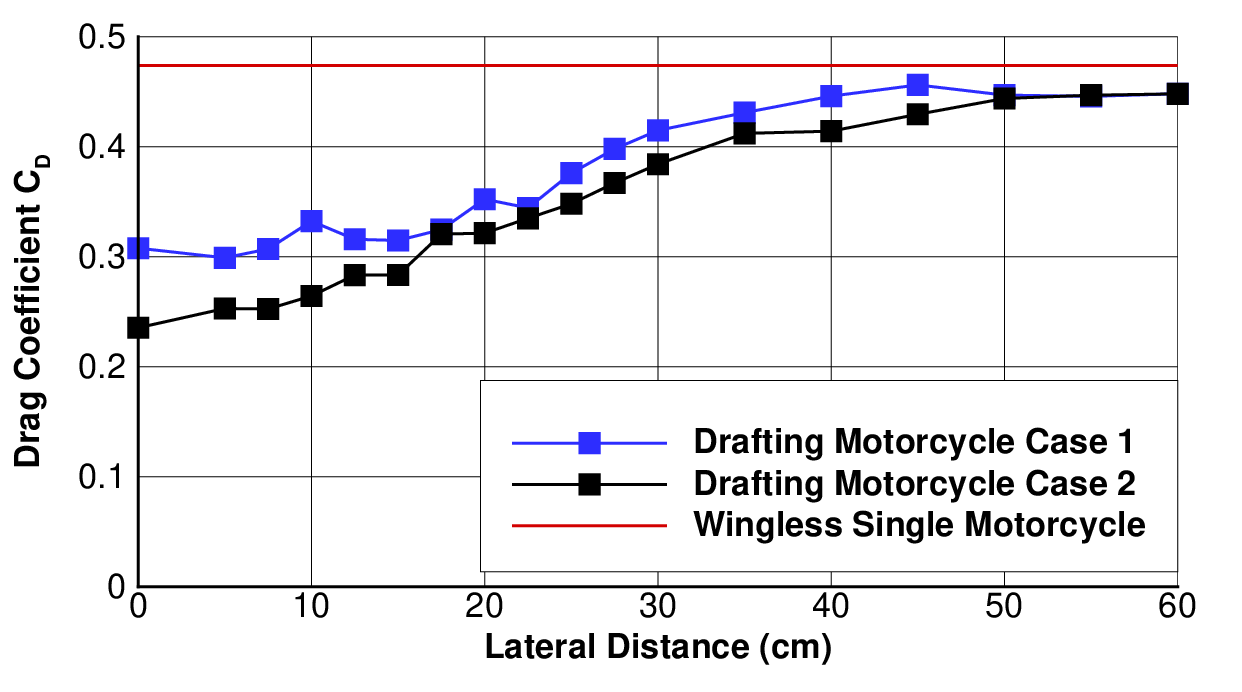}
\includegraphics[width=0.45\columnwidth]{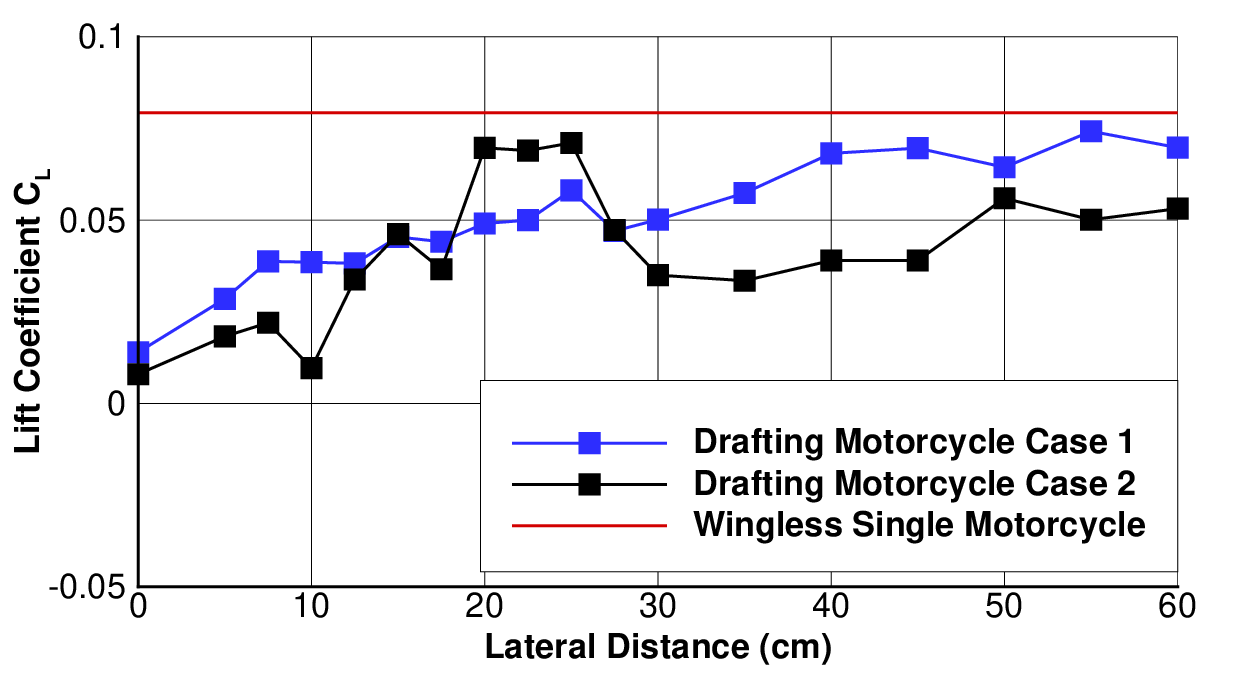} \\
\includegraphics[width=0.45\columnwidth]{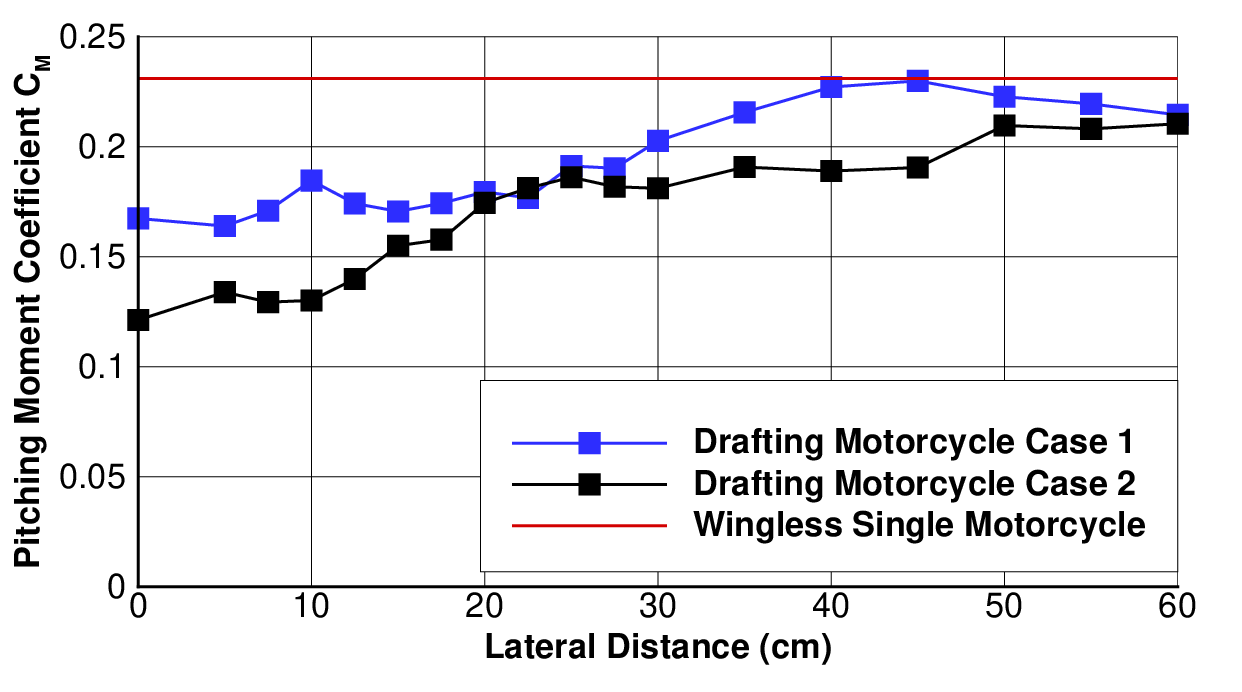}
\caption{Aerodynamic coefficients variation in function of lateral distance for the front motorcycle without wings (case 1) and with wings (case 2).}
\label{fig_13}
\end{figure}

Figure \ref{fig_13} shows the influence of the lateral distance variation over the three main aerodynamic coefficients. Reduction for all the aerodynamic coefficients can be noted for small values of the lateral position variation between both motorcycles due to the increased drafting effect already observed in figure \ref{fig_12}. At the interval ranging from approximately $y=20$ to $y=25$ cm the lift coefficient in case 2 is larger than in case 1 and then retains smaller values for larger values of lateral distance. This increase in the lift coefficient is due to the trailing motorcycle being aligned with the induced wingtip vortex core that is convected downstream, where complex aerodynamic interaction occurs. Also, a reduction of the lift coefficient ranging from $y=25$ cm to larger lateral distances can be noted, where the downwash velocity component induced by the wingtip vortex produces beneficial effects over the trailing motorcycle from a security and stability point of view, resulting into a smaller wheeling tendency. The drag and pitching moment coefficients retain smaller values for case 2 when compared to case 1 and in some points near values as well.



\subsection{Case 3: Aerodynamic Effect on Both Wing-Equipped Motorcycles}

The last case refers to the aerodynamic influence of the wing-equipped leading motorcycle over the pursuing one also equipped with wings on its fairings. This is the case found nowadays in the MotoGP races since all the teams use motorcycles with aerodynamic attachments for downforce generation. The same longitudinal and lateral position variations are assessed as in cases 1 and 2. The aerodynamic coefficients are presented for both wing-equipped motorcycles and as a matter of comparison, the wingless trailing motorcycle coefficients are also presented to show the behavior reflecting changes in flow regime.

\begin{figure}[ht!]
\centering
\includegraphics[width=0.45\columnwidth]{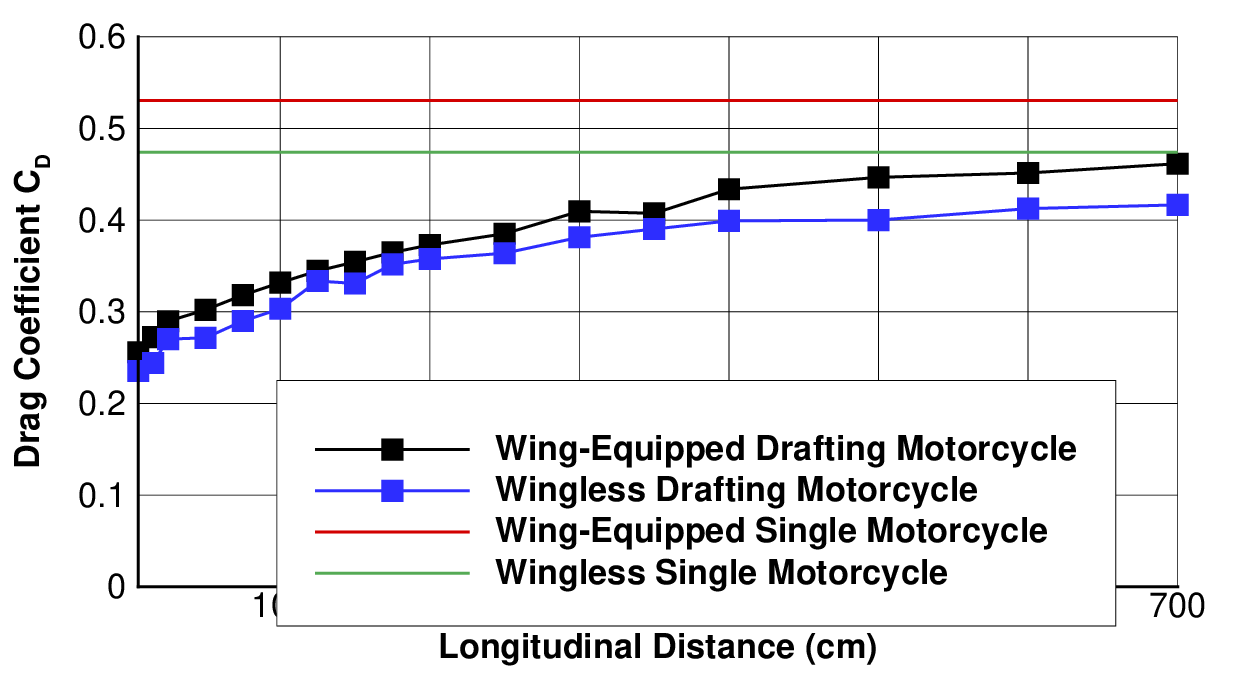}
\includegraphics[width=0.45\columnwidth]{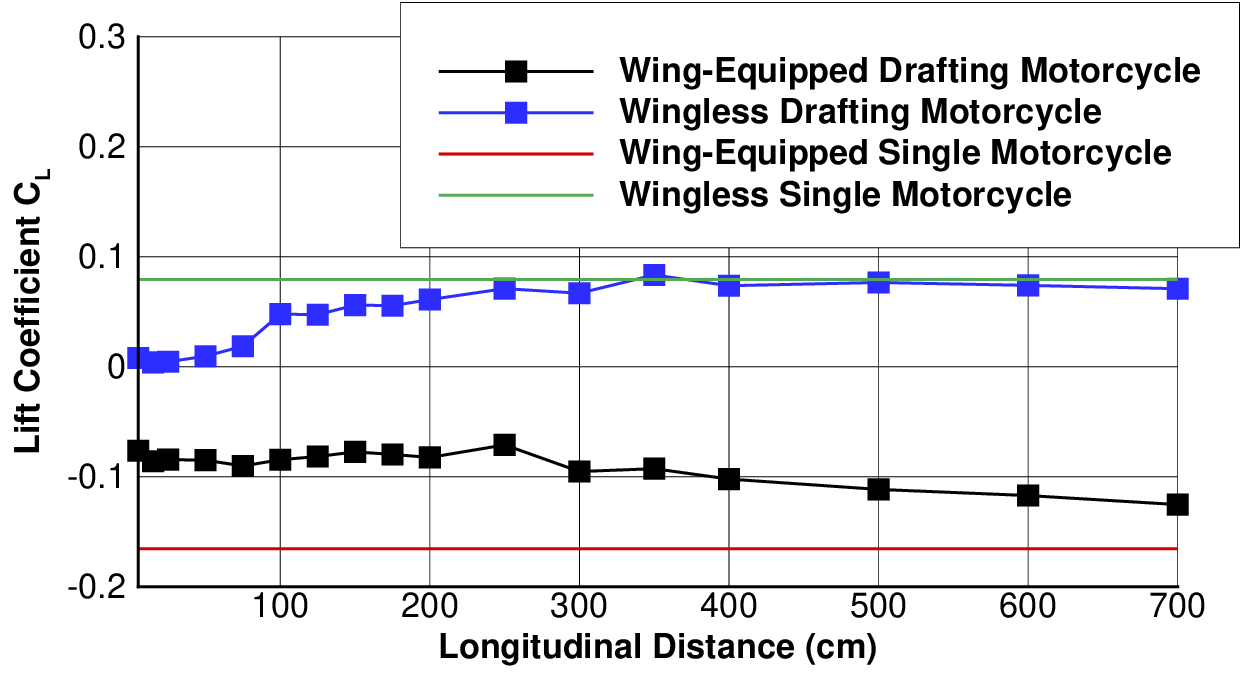} \\
\includegraphics[width=0.45\columnwidth]{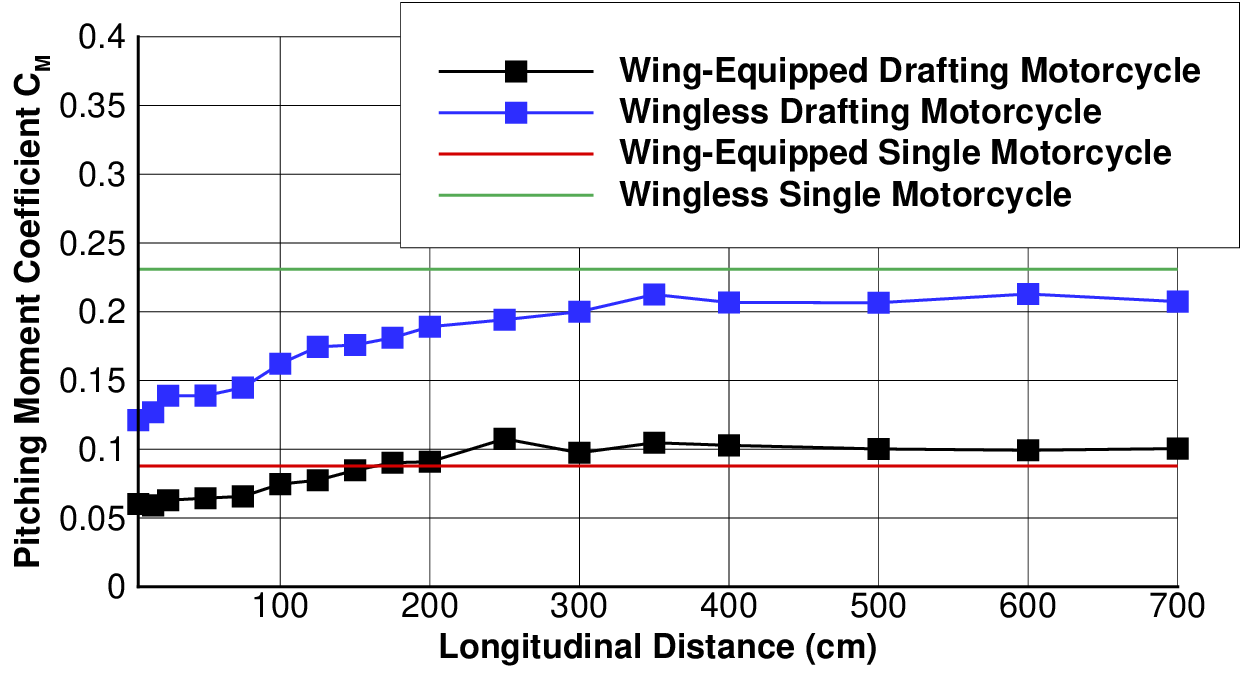}
\caption{Aerodynamic coefficients variation due to longitudinal distance between the leading and the drafting motorcycle of case 3.}
\label{fig_15}
\end{figure}

Figure \ref{fig_15} shows the variation of the aerodynamic coefficients as a function of the longitudinal distance between motorcycles. As expected, due to the induced drag generated by the downforce, the wing-equipped motorcycle produces more drag than the wingless one. The total drag difference between the two configurations grows as the drafting effect decreases, resulting on a rise on the front wing effectiveness and the recovery of their respective leading motorcycle base drag coefficient.

Also in figure \ref{fig_15}, the lift coefficient shows small changes with an increase in longitudinal distance, although the generated downforce remains roughly half of the generated value for the leading motorcycle. This affects entirely the braking performance of the drafting motorcycle and can be regarded as one of the main hypotheses for the lack of braking capacity in race conditions, along with the drag reduction due to drafting. The downforce is regained as the longitudinal distance grows larger for the wing-equipped motorcycle and the one with wingless configuration also regains its undisturbed flow lift coefficient.

The pitching moment coefficient in figure \ref{fig_15} shows almost no change when longitudinal distance variation is considered, where a small reduction can be observed for closer distances, while the coefficient of the undisturbed flow is recovered for short distances as small as $x=200$ cm. close proximity results in a reduced relative pitching moment due to the low-pressure wake impinging over the pursuing motorcycle, where the overall body format produces the positive values. The same trend can be noted for both configurations, while the front wings reduce the wheeling tendency for all longitudinal distance values.

\begin{figure}[ht!]
\centering
\includegraphics[width=0.45\columnwidth]{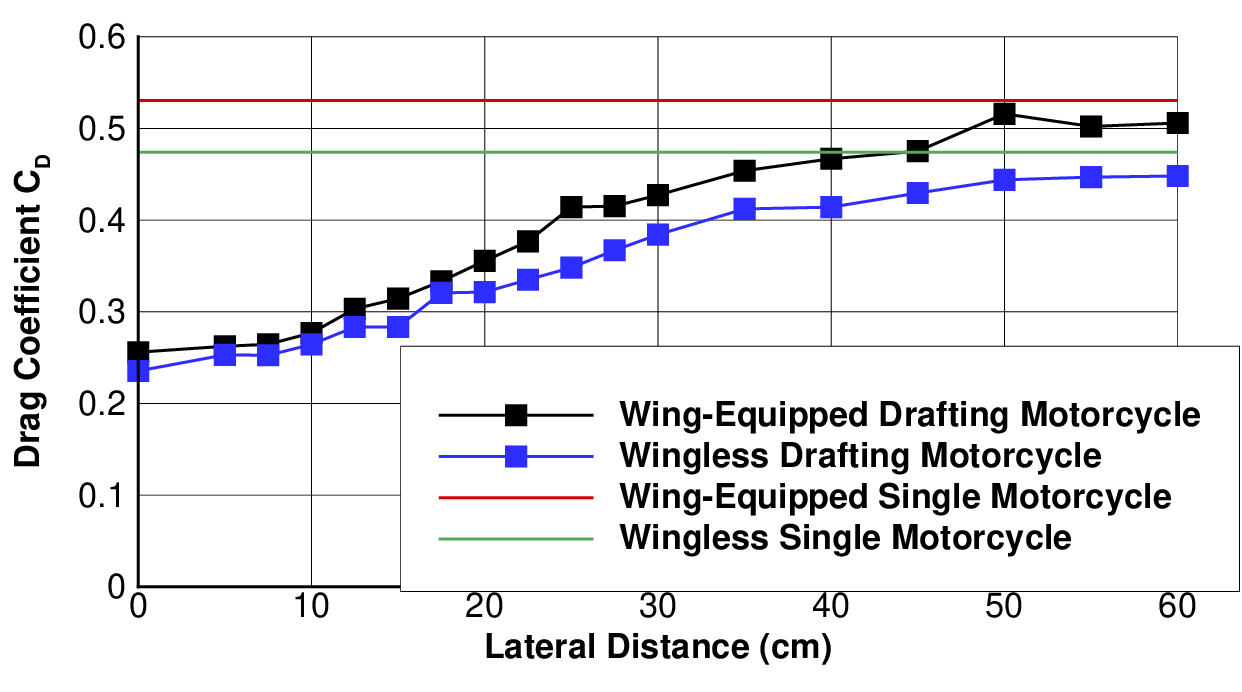}
\includegraphics[width=0.45\columnwidth]{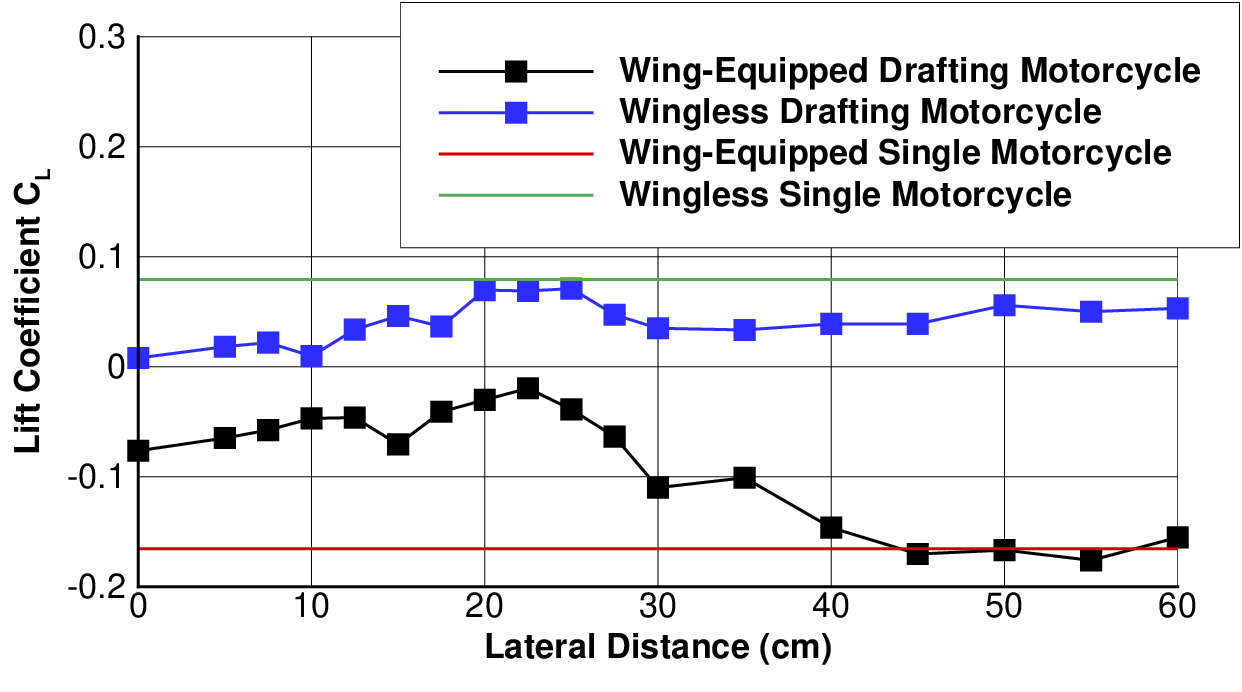} \\
\includegraphics[width=0.45\columnwidth]{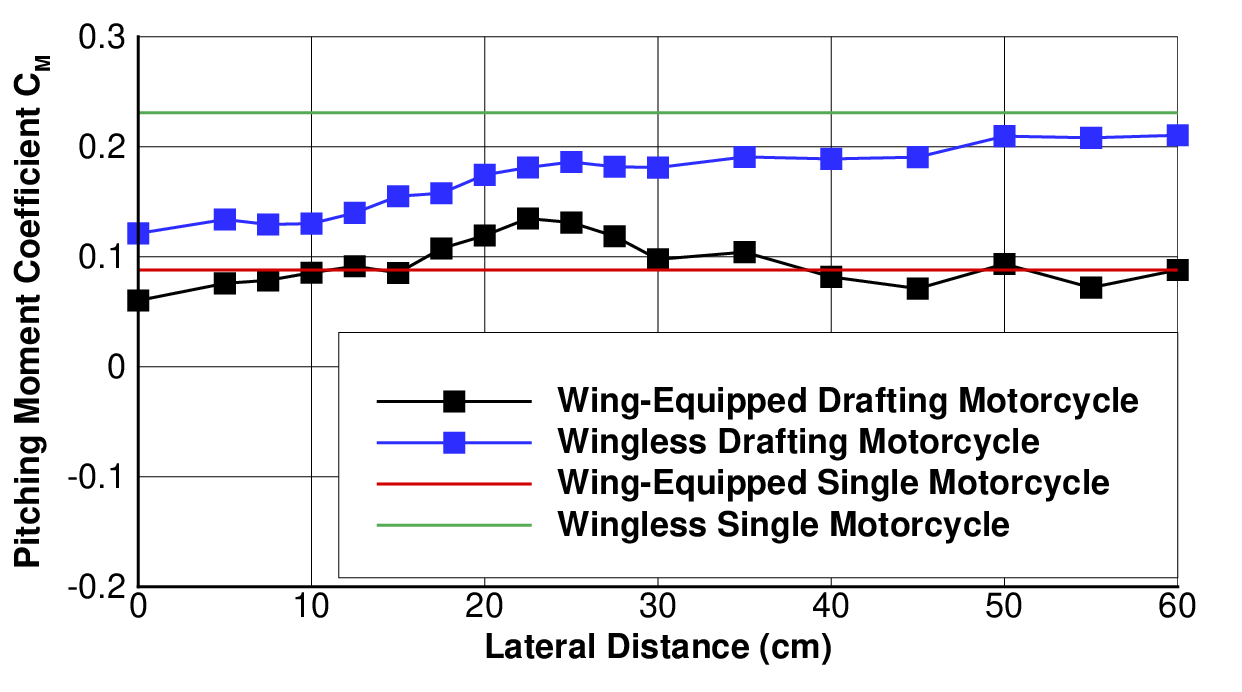}
\caption{Aerodynamic coefficients variation due to lateral distance between the leading and the drafting motorcycle of case 3.}
\label{fig_16}
\end{figure}

Figure \ref{fig_16} shows the aerodynamic coefficients as a function of the lateral distance variation between the leading and the drafting motorcycle. The drag force follows the same trend as the case of the wingless pursuing motorcycle, while overall drag is almost recovered when compared to the motorcycle without incoming flow disturbance.

The generated downforce as a function of the lateral distance shows some oscillation due to the upwash induced velocity component and the convected vortex core for small values of lateral distance, where a near zero value is achieved. For a larger lateral deviation the obtained downforce value is similar to the reference coefficient value of the single wing-equipped motorcycle. The similar value is due to the downwash velocity component raising the effective angle of attack imposed at the wings, while the wingless pursuing motorcycle configuration retains a positive valued lift for all the lateral distances.

The pitching coefficient of the trailing wing-equipped motorcycle shows no greater change as a function of the lateral deviation. For small values of lateral distance, the upwash still has no effect due to the close longitudinal proximity between the motorcycles, while at larger values the position where the vortex core is aligned to the drafting motorcycle has little effect over the wheeling tendency. At larger values of lateral distance, the pitching moment is practically the same as the reference value of the case of undisturbed incoming flow. The wingless motorcycle configuration shows a steady increase in the pitching moment, attaining the undisturbed reference value as given by table \ref{tab_02}.

\section{Conclusion}

This paper dealt with the aerodynamic influence of the leading motorcycle over a drafting one in race conditions for the sports bike competitions, where typical examples are the MotoGP racing prototypes and the WSBK production based categories. In the assessed cases there are wing-equipped motorcycles that may influence the pursuing counterparts in mixed ways, aside from the well-known drafting effect.

It was identified that the coherent vortex generated by the wingtip used for downforce generation acts as a modifier of the effective angle of attack in the same manner as a following aircraft and where its effects may affect negatively or may also give a competitive advantage for the pursuing motorcycle depending on its relative position. A base case of both motorcycles with no wings was established as a case for comparison and it was found that all the listed aerodynamic coefficients suffered reduction as a function of the longitudinal and lateral distances. The drafting effect is the sole responsible for this reduction, since no coherent flow structure is generated by a leading wingless motorcycle, aside from the usual low-pressure, chaotic and turbulent wake akin to one from a blunt body.

The front motorcycle with a downforce generation wing fitted on its front fairing affected positively the pursuing motorcycle for very close longitudinal distances while the drag remained almost the same for all the distance range. A rise in the lift and pitching moment coefficients for larger distances was observed as well, bringing negative effects related to the stability and thus safety to the pursuing motorcycle. It is important to note that this relative position with the negative effects is the usual one during motorcycle races, where packs of lining motorcycles run for better draft and race lines. The lateral variation somehow resulted in better aerodynamic performance for the pursuing motorcycle, taking advantage of the aerodynamic benefits from the downwash velocity component generated by the wing of the leading motorcycle. The last stage of the overtaking process is the one where a lateral distance deviation takes place between motorcycles before the turn approximation and braking stages.

\begin{figure}[ht!]
\centering
\includegraphics[width=0.95\columnwidth]{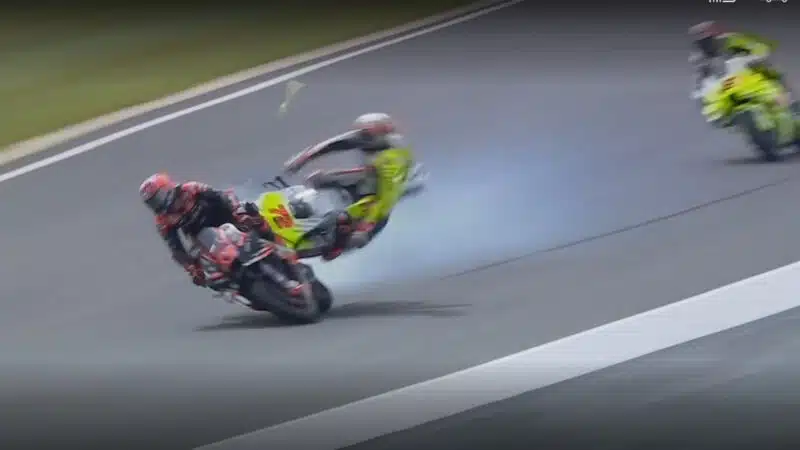}
\caption{Maverick Viñales and Marco Bezzecchi crash due to strong aerodynamic influence during the 2024 Australian Motorcycle Grand Prix sprint race \cite{vinales_bezz_crash}.}
\label{fig_20}
\end{figure}

The effect of downforce generation wings fitted on the pursuing motorcycle is mostly positive and reduces up to a certain extent the adverse effects of the downstream convected induced flow by the leading one. Nevertheless, the same induced flow represents dangerous conditions for racing as a whole, affecting critical stages around the racetrack. One fine but dangerous example is the crash between Maverick Vin\~ales and Marco Bezzecchi during the sprint race of the 2024 Australian Motorcycle Grand Prix, where Bezzecchi could not brake properly after being overtaken by Viñales, due to not having enough aerodynamic drag and downforce, as depicted in figure \ref{fig_20}.

Motorcycle racing has been one of the most dangerous kinds of motorsport since its dawn at the beginning of the $20$th century. Paradoxically it has never been safer due to the better-performing brakes, chassis, tires and safety equipment. While on the quest for better racing performance, the introduction of aerodynamic appendices brought lap time reduction but also less stability for pursuing motorcycles at several lap stages. A sudden complete removal of the front wings could bring several performance and balance issues to the current racing prototypes and the production motorcycles as well, since their design process takes the full package into consideration. A full commitment of the racing governing bodies must consider the safety issues from such technology, taking into account their influence on the sport as a whole, not only the competition issues. A steady reduction of the allowed wing geometry is underway for the 2027 MotoGP season, but further reduction and possibly full elimination of these aerodynamic appendices must be considered, as suggested by the results shown in this work.

\printbibliography


%




\end{document}